\documentclass[aps,twocolumn,floatfix]{revtex4}
\usepackage{times}
\usepackage{graphics}
\usepackage{epsfig,calc}

\usepackage{bm}

\usepackage{amsmath,amssymb,chemarr}

\def\btt1{{\tt$\backslash$\string1}}%

\def\AmS{{\protect\the\textfont2
        A\kern-.1667em\lower.5ex\hbox{M}\kern-.125emS}}

\citeindextrue

\newcommand{\kt}{k_{\text{B}}T}
\newcommand{\ec}{\varepsilon_{\text{c}}}
\newcommand{\eb}{\varepsilon_{\text{b}}}
\newcommand{\fsp}{\hat{k}_\text{as}}
\newcommand{\bsp}{\hat{k}_\text{b}}
\newcommand{\gh}{g_{\text{H}}}
\newcommand{\gb}{g_{\text{b}}}
\newcommand{\Rc}{R_\text{C}}
\newcommand{\Rcap}{R_\text{cap}}
\newcommand{\sigc}{\sigma_\text{C}}

\newcommand{\qs}{q_\text{p}}
\newcommand{\CS}{\rho_\text{p}}
\newcommand{\CB}{C_\text{p}}
\newcommand{\CC}{\rho_\text{C}}
\newcommand{\CCmolar}{C_\text{C}}
\newcommand{\npratio}{r_\text{CP:NP}}
\newcommand{\ncontact}{n^\text{c}}

\newcommand{\ssb}{s_\text{rot}}
\newcommand{\sdegen}{S^{\text{degen}}}
\newcommand{\Pnm}{P_{n,m}}
\newcommand{\Gcore}{G^\text{core}}
\newcommand{\Gcap}{G^\text{cap}}
\newcommand{\Gcs}{G^\text{surf}}
\newcommand{\Smix}{S^\text{mix}}
\newcommand{\FSF}{f_\text{SF}}
\newcommand{\Fsf}{\FSF}
\newcommand{\FSFEC}{f_\text{SF,EC}}

\newcommand{\Fec}{F_\text{EC}}
\newcommand{\Fex}{f_\text{EX}}
\newcommand{\Fcore}{F_\text{core}}
\newcommand{\Sad}{S^\text{ad}}
\newcommand{\Ac}{A_\text{C}}

\newcommand{\tmax}{\theta_{\text{m}}}
\newcommand{\pmax}{\phi_{\text{m}}}
\newcommand{\zg}{z_\text{p}}
\newcommand{\zc}{z_\text{TEG}}

\newcommand{\kad}{k_\text{ad}}
\newcommand{\sbar}{\bar{s}}
\newcommand{\nnuc}{n_\text{nuc}}
\newcommand{\zp}{z^\prime}
\newcommand{\ip}{i^\prime}
\newcommand{\up}{u^\prime}
\newcommand{\phib}{\phi^\text{b}}
\newcommand{\phiWater}{\phi_{\text{H}_2\text{O}}}
\newcommand{\phiHplus}{\phi_{\text{H}_3\text{O}^{+}}}
\newcommand{\as}{a_\text{s}}

\newcommand{\srf}{k_\text{as}}
\newcommand{\SRF}{\srf}

\newcommand{\pka}{pK_\text{a}}
\newcommand{\pH}{p\text{H}}
\newcommand{\csurf}{\rho^\text{eq}_\text{surf}}
\newcommand{\rsurf}{\rho_\text{surf}}
\newcommand{\sg}{\sigma_g}
\newcommand{\molsec}{\mbox{M s}^{-1}}

\newcommand{\w}{\overline{W}}
\newcommand{\om}{\overline{\Omega}}
\newcommand{\Wspont}{\rho_1 W^\text{spont}}
\newcommand{\Wsp}{W^\text{spont}}
\newcommand{\Wspontbar}{\overline{W}^\text{spont}}

\begin{document}
\baselineskip=1.3\baselineskip

\title{A theory for viral capsid assembly around electrostatic cores}
\author{Michael F. Hagan}
\affiliation{Department of Physics, Brandeis University, Waltham, MA, 02454}
\date{\today}
  \begin{abstract}We develop equilibrium and kinetic theories that describe the assembly of viral capsid proteins on a charged central core, as seen in recent experiments in which brome mosaic virus (BMV) capsids assemble around nanoparticles functionalized with polyelectrolyte.  We model interactions between capsid proteins and nanoparticle surfaces as the interaction of polyelectrolyte brushes with opposite charge, using the nonlinear Poisson Boltzmann equation.  The models predict that there is a threshold density of functionalized charge, above which capsids efficiently assemble around nanoparticles, and that light scatter intensity increases rapidly at early times, without the lag phase characteristic of empty capsid assembly. These predictions are consistent with, and enable interpretation of, preliminary experimental data.  However, the models predict a stronger dependence of nanoparticle incorporation efficiency on functionalized charge density than measured in experiments, and do not completely capture a logarithmic growth phase seen in experimental light scatter. These discrepancies may suggest the presence of metastable disordered states in the experimental system.  In addition to discussing future experiments for nanoparticle-capsid systems, we discuss broader implications for understanding assembly around charged cores such as nucleic acids.
\end{abstract}

 \maketitle

\section{Introduction}
 \label{sec:intro}

The cooperative assembly of heterogenius building blocks into ordered structures is crucial for many biological processes, such as the assembly of protein subunits around a viral nucleic acid to form a capsid.  Understanding how nucleic acid-protein interactions drive cooperative assembly could enable antiviral drugs that block this essential step in viral replication.  At the same time, engineered structures in which viral capsids assemble around synthetic cargoes show great promise as delivery vehicles for drugs \cite{Gupta2005,Garcea2004,Dietz2004} or imaging agents \cite{Soto2006,Sapsford2006,Boldogkoi2004,Dragnea2003} and as subunits or templates for the synthesis of advanced multicomponent nanomaterials \cite{Chatterji2005,Falkner2005,Flynn2003,Douglas1998,Douglas2006}. Realizing these goals, however, requires understanding at a fundamental level how properties of cargoes and capsid proteins control assembly rates and mechanisms. With that goal in mind, this article presents coarse-grained thermodynamic and kinetic models that describe experiments in which brome mosaic virus (BMV) capsid proteins assemble around spherical charge-functionalized nanoparticle cores.

{\bf Identifying mechanisms for simultaneous assembly and cargo-encapsidation.}
Elegant experiments have studied the assembly of capsids around central cores consisting of nucleic acids (e.g. \cite{Johnson2004, Fox1994, Valegard1997, Johnson2004b,Tihova2004,Krol1999,Stockley2007,Toropova2008}, inorganic polyelectrolytes \cite{Bancroft1969,Hu2008,Sikkema2007}, and charge-functionalized cores \cite{Sun2007,Dixit2006,Chen2005,Dragnea2003,Chen2006,Chang2008}.  However, assembly mechanisms of viruses and virus-like particles remain incompletely understood because cargo-protein interactions and the structures of transient assembly intermediates are not accessible to experiments.   For the case of empty-capsid assembly, Zlotnick and coworkers achieved great insight about assembly mechanisms with simplified  theories (e.g.~\cite{Zlotnick2007,Zlotnick1994,Endres2002,Zlotnick2000,Ceres2002,Singh2003,Kegel2004}), for which a small number of parameters are fit by comparison with experimental assembly data over a range of protein concentrations and pH values.  The theories developed in this work enable similar comparisons for the assembly of capsids around electrostatic cores, and could thereby elucidate mechanisms for simultaneous assembly and cargo encapsidation.

Prior theoretical works have led to important insights about assembly around polymeric cores, but were equilibrium studies \cite{Belyi2006,Angelescu2006,Schoot2005, Zhang2004,Zandi2008,Hu2007a} or considered specific RNA-mediated assembly pathways with phenomenological descriptions of protein-nucleic acid interactions \cite{Rudnick2005,Hu2007b}.  Computer simulations of a model in which subunits assemble around rigid spherical cores \cite{Elrad2008,Hagan2008} suggest that a core with a geometry commensurate with the lowest free energy empty-capsid morphology can increase assembly rates and the efficiency of assembly (compared to empty capsid assembly\cite{Hagan2006,Rapaport1999,Rapaport2004,Rapaport2008,Nguyen2007,Zhang2006,Sweeney2008,Schwartz1998,Wilber2007,Nguyen2008}) through the effects of heterogeneous nucleation and templating; however, assembly is frustrated for core-subunit interaction strengths that are too strong\cite{Elrad2008,Nguyen2008b}.  Cores that are not size-matched with the lowest free energy capsid morphology can direct assembly into alternative capsid morphologies for optimal core-subunit interaction strengths \cite{Elrad2008}.   Important limitations of the simulations are that extensive comparison with experimental data is computationally intractable, and the simulated models do not explicitly represent electrostatics or the polymeric nature of disordered N-terminal tails in capsid proteins.

  The present work aims to overcome these limitations by calculating the electrostatic interaction between charge-functionalized nanoparticles and capsid proteins, and then constructing simplified thermodynamic and kinetic theories that describe the simultaneous assembly and cargo encapsidation process.    We explore the effect of cargo-capsid protein interactions on assembly mechanisms by presenting predicted assembly rates and nanoparticle incorporation efficiencies as functions of the functionalized charge density, nanoparticle-protein stoichiometric ratio, and the capsid protein concentration. Because each of these parameters can be experimentally controlled, the theoretical predictions suggest a series of experiments that can elucidate the relationship between cargo-capsid protein interactions and assembly mechanisms.

\begin{figure} [bt]
\epsfig{file=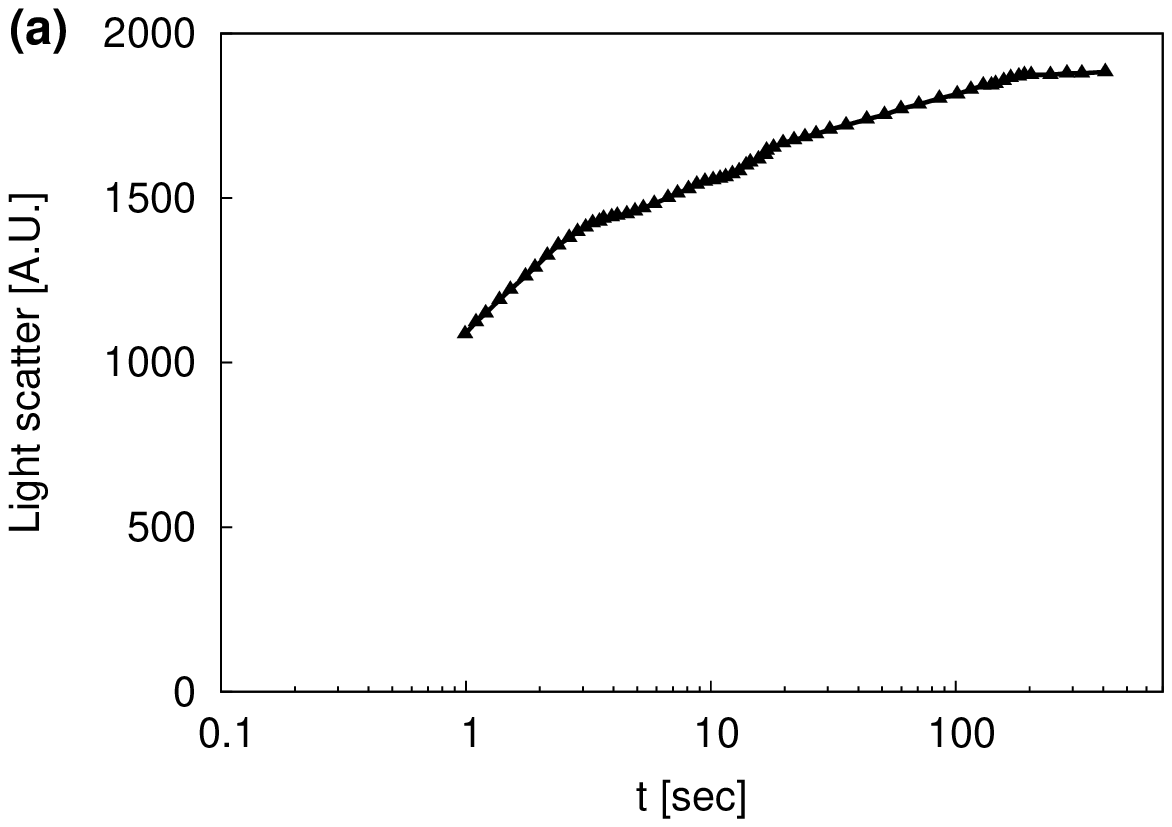,width = \textwidth/2}
\epsfig{file=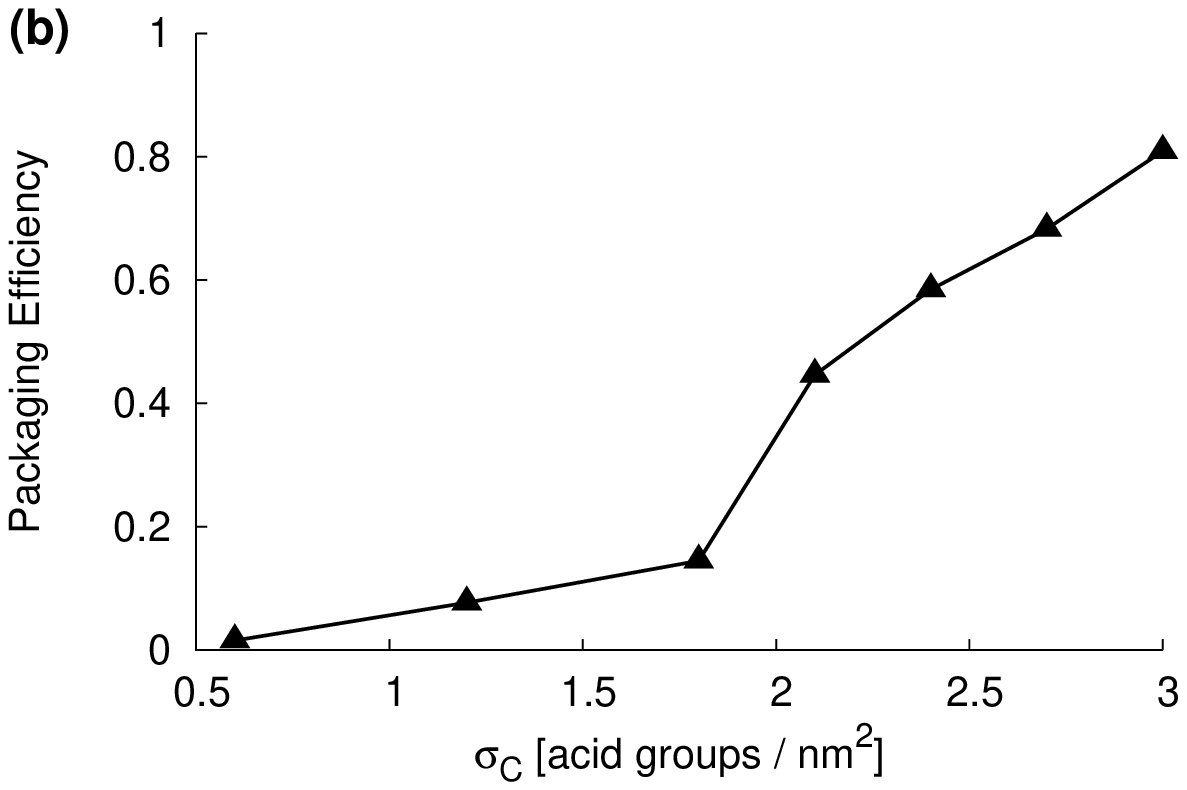,width = \textwidth/2}
\caption{\label{fig:experimentalData}
{\bf(a)} Time-resolved light scatter for capsid assembly on nanoparticles with a functionalized surface density of $\sigc=3$ acid groups/nm$^2$.   {\bf (b)} Packaging efficiency, or the fraction of nanoparticles incorporated into $T$=3 capsids (measured from TEM microgaphs), varies with $\sigc$.     The capsid protein-nanoparticle stoichiometric ratio is $\npratio=1.33$, core  diameters are 12 nm (which promote $T$=3 capsid formation), the capsid protein concentration is 0.4 mg/ml ($\CB=10$ $\mu$M dimer subunit), and there is 100 mM 1:1 salt and 5 mM 2:1 salt.  Packaging efficiencies were measured at $t=10^5$ seconds.  The surface density of acid groups is estimated from the fraction of carboxylated TEG molecules, assuming a total surface density of 3 TEG molecules/nm$^2$ on the nanoparticle surface. Data thanks to Bogdan Dragnea \cite{Dragnea2008}.}
\end{figure}

Although there is currently not sufficient experimental data to estimate values for several unknown parameters (adsorption and assembly rate constants, and subunit-subunit binding energies), we qualitatively compare the theoretical predictions to preliminary experimental time-resolved light scattering data (Fig.~\ref{fig:experimentalData}a) and measurements of the efficiency of nanoparticle incorporation in well-formed capsids (Fig.~\ref{fig:experimentalData}b).  With physically reasonable values for the unknown parameters, the predicted light scatter increases rapidly at short times without a lag phase, consistent with the experimental measurements, but the theory does not completely capture the functional form of light scatter at long times.  The theory predicts that nanoparticles are incorporated into capsids only above a threshold functionalized charge density, which is similar to the lowest charge density for which significant incorporation is observed in experiments. However, predicted incorporation efficiencies rise to 100\% over a narrow range of charge densities, whereas the experimental data shows a gradual increase in the incorporation efficiency with increasing charge density. We discuss experiments and potential improvements to the theory which might elucidate these discrepancies.

  This paper is organized as follows.  In Section \ref{sec:two} we calculate the interaction between capsid proteins and functionalized charge on nanoparticle surfaces.  In Section \ref{sec:three}, we present a model for the thermodynamics of capsid assembly in the presence of rigid spherical cores, followed in Section \ref{sec:kineticTheory} by a kinetic theory that describes the simultaneous assembly of capsid subunits on nanoparticle surfaces and in bulk solution.  We analyze kinetics and the assembly pathways predicted by the model in section \ref{sec:ktResults}, and discuss the connection to current and future experiments in  Section~\ref{sec:discussion}.  We also compare theory predictions to simulation results in Appendix A.

{\bf Experimental motivation.} This work is motivated by experiments in which BMV capsid proteins assemble around gold nanoparticle cores \cite{Chen2006,Sun2007} functionalized with thiolalkylated tetraethylene glycol (TEG) molecules, a fraction of which are terminated with carboxylate groups \cite{Chen2006}.  With 12 nm nanoparticles the TEG-nanoparticle complexes have a diameter of roughly 16 nm, which is nearly commensurate with the interior of a T=3 BMV capsid (17-18 nm \cite{Lucas2002}). Ionized carboxyl groups provide a driving force for capsid proteins to adsorb and assemble on nanoparticle surfaces.
Time-resolved light scatter measurements (Fig.~\ref{fig:experimentalData}a), which monitor assembly kinetics, are strikingly different than light scatter of empty capsid assembly at short times (e.g. \cite{Chen2008,Zlotnick1999,Casini2004}); instead of a lag phase ($\sim 10$ seconds -- minutes) there is a rapid ($\sim 1$ second) rise in signal.  As described in Section~\ref{sec:ktResults}, the theoretical predictions in this work begin to explain this observation.  Assembly effectiveness is monitored by using TEM to measure packaging efficiencies, or the fraction of nanoparticles incorporated within well-formed capsids. As shown in Fig.~\ref{fig:experimentalData}b, packaging efficiencies increase from 0\% to nearly 85\% as the fraction of TEG molecules that are carboxylate-terminated increases from 20\% to 100\%.

While the light scattering data was obtained from an assembly reaction entirely at $\pH$ = 5, the packaging efficiency data shown in Fig.~\ref{fig:experimentalData}a were obtained with a protocol in which capsid subunits were first dialyzed against a buffer at $\pH = 7.4$, followed by buffer at $\pH = 5$. In this work we consider only $\pH=5$, since at $\pH = 7.4$ disordered protein-nanoparticle complexes form but well-formed capsids are not observed without subsequently lowering $\pH$. Assembly may not be favorable at $\pH = 7.4$ because binding interactions between capsid proteins are less favorable than at lower $\pH$ \cite{Ceres2002}; also strong TEG-protein interactions might render disordered states kinetically or thermodynamically favorable.

\begin{figure} [bt]
\epsfig{file=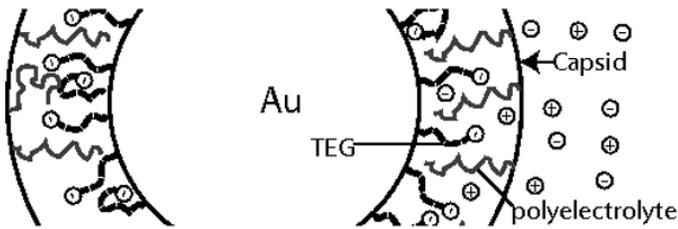,width=\textwidth/2}
\caption{\label{fig:coreCapsidGeometry}
The model system. Surface functionalization molecules (TEG) are end-grafted to an impenetrable gold sphere. The cationic capsid protein N-terminal arms are modeled as polyelectrolytes grafted to the inner surface of a sphere (the capsid), which is impermeable to TEG and polyelectrolyte but permeable to solvent and ions.}
\end{figure}

\begin{figure} [bt]
\epsfig{file=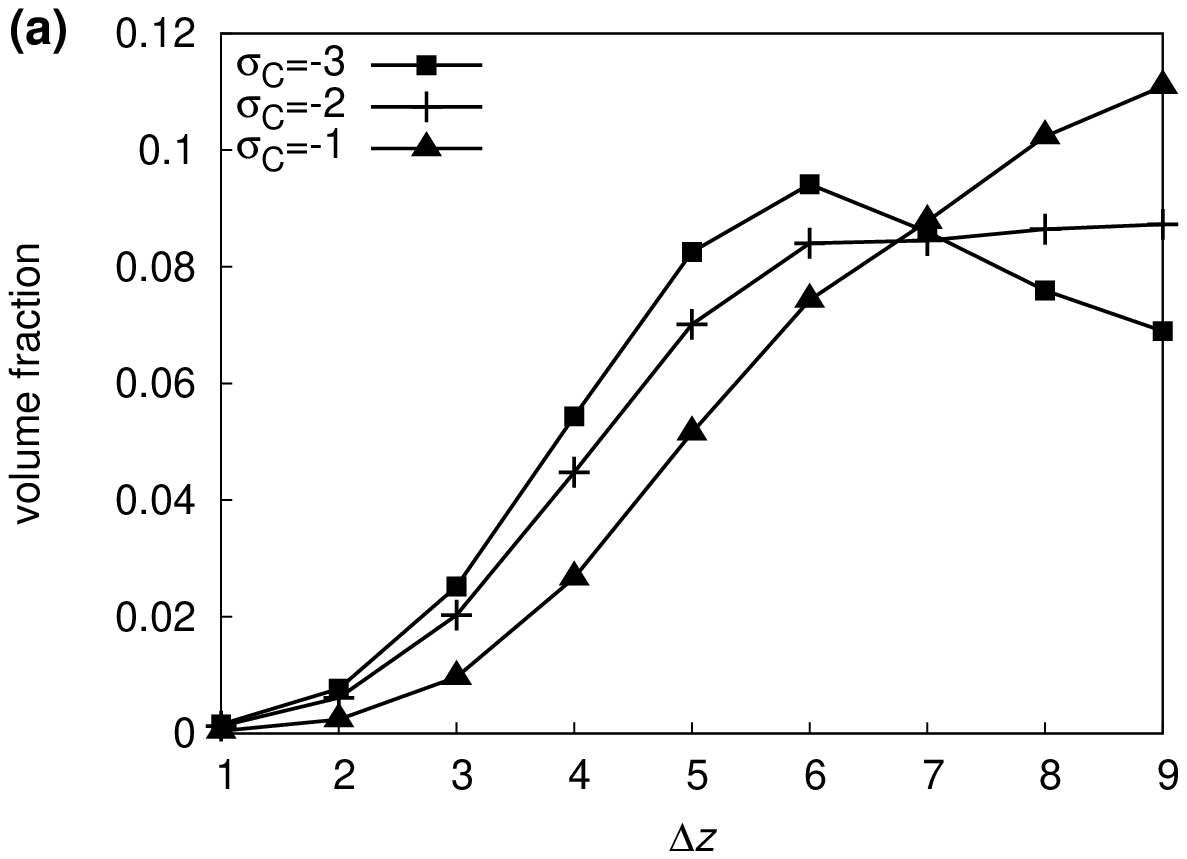,width=\textwidth/2}
\epsfig{file=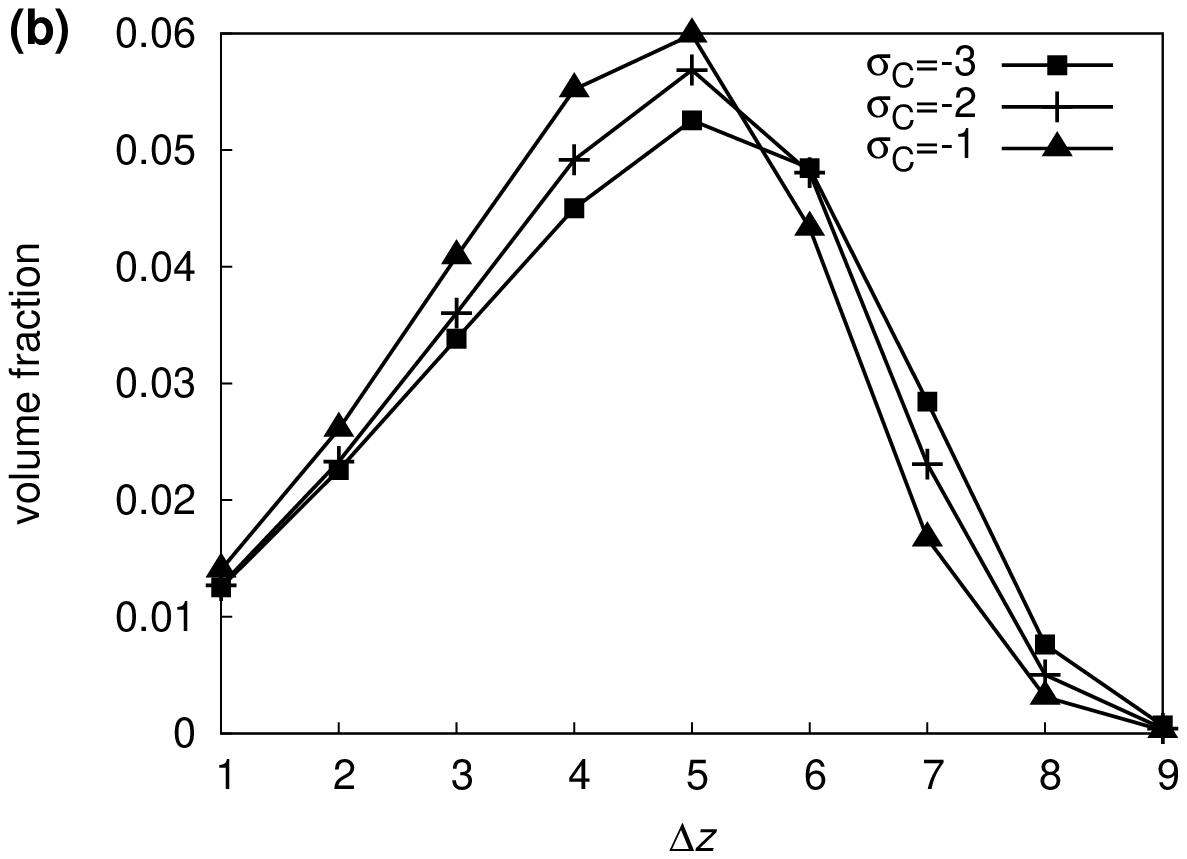,width=\textwidth/2}
\caption{\label{fig:layerHeights}
  The volume fractions of {\bf (a)} polyelectrolyte segments and {\bf (b)} carboxylate groups in the grafted TEG layer are shown as a function of layer ($\Delta z=z-\Rc/l+1$) above the nanoparticle surface for several surface acid group densities, with the nanoparticle radius $\Rc=6$ nm and the layer dimension $l=0.35$ nm.}
\end{figure}

\section{Subunit adsorption on nanoparticle surfaces}
\label{sec:two}
For many capsid proteins, there is a high density of basic residues on flexible N-terminal tails, with numbers of charged amino acids that range from six to 30 \cite{Belyi2006}.  These positive charges help drive capsid assembly \emph{in vivo} by interacting with negative charges on nucleic acids, and the interaction of polymeric tails and nucleic acids is explored in Refs.~\cite{Belyi2006,Hu2007b,Siber2007}.  In this section, we analyze the interactions between capsid protein tails and surface functionalized TEG molecules.  For simplicity, we neglect other effects, such as hydrophobic interactions, that might drive subunit adsorption, and we assume that all charges on capsid proteins are located in the polymeric tails. For the remainder of this work, we will reserve the word `polyelectrolyte' to apply to the charged polymeric tails on capsid proteins; we will not use this word to refer to the functionalized TEG molecules.

{\bf Model system.}  We first consider a dilute solution of capsid protein subunits with density $\CS$, and a single nanoparticle; we consider a finite concentration of cores in Section \ref{sec:three}.  At high surface coverage or in an assembled capsid the folded portion of capsid proteins acts as an impenetrable barrier to the polymeric tails.  We therefore represent the nanoparticle-capsid complex as a concave polyelectrolyte brush (the interior surface of the capsid) interacting with a convex brush of the opposite charge (the TEG layer); the model system geometry is illustrated in  Fig.~\ref{fig:coreCapsidGeometry}.  We consider surface functionalization molecules grafted to the outer surface of an impenetrable sphere with radius 6 nm. The TEG surface functionalization molecules used in Ref.~\cite{Sun2007} consist of a 10 carbon alkanethiol chemically linked to a polyethylene glycol chain of 4 monomer units.  The surface density of functionalized charge groups, $\sigc$, is experimentally controlled by terminating a fraction TEG molecules with carboxylates, with the remainder terminated with neutral groups \cite{Dragnea2008}. We represent TEG molecules as freely jointed chains  with 9 neutral segments end-grafted to the surface and one charged segment (with variable valence) at the free end.

 The N-terminal tails, or 'arms' of BMV capsid proteins contain nine positive charges in the first 20 amino acids and the first 25-40 residues are disordered in capsid crystal structures \cite{Lucas2002}.  Since  BMV capsids assemble from protein dimer subunits \cite{Pfeiffer1974,Cuillel1983,Berthetcolominas1987}, in our model each subunit has a net charge of $\qs=18$.  We represent the two N-terminal arms from each dimer subunit as a single polymer of 36 segments with Kuhn length 3.5 $\AA$, and valence 0.5. These polyelectrolytes are end-grafted to the inner surface of a sphere with radius 9 nm, the inner diameter of a T=3 BMV capsid \cite{Lucas2002}.  The model capsid is permeable to solvent and ions, but impenetrable to TEG and polyelectrolytes segments. The latter condition is reasonable for high surface coverages of capsid subunits and in the strong coupling limit, in which case the polyelectrolyte configurational entropy is negligible compared to electrostatic effects. For low charge densities, the results do not change qualitatively if the impermeable capsid is eliminated, but the equilibrium concentration of adsorbed polyelectrolytes increases by up to 50\%.

 \emph{Numerical calculations.}
To our knowledge, interacting polyelectrolyte brushes with this geometry have not been studied, although the interaction of capsid arms with a nucleic acid is considered in Refs.~\cite{Belyi2006,Siber2008}.  For many cases we consider, the electrostatic potential in the region of the capsid and nanoparticle surface is large; therefore, we will solve the nonlinear Poisson Boltzmann (PB) equation.
To do so, we employ the method of Scheutjens and Fleer (SF) \cite{Scheutjens1979,Bohmer1990,Bohmer1991}, in which the spatial distributions of polyelectrolyte, TEG, ions, and water, as well as the dissociation equilibrium for carboxylate and water groups, are solved numerically with a self consistent field approximation on a lattice.  The calculation accounts for the finite size of all species and the calculated free energies explicitly consider the entropy of ions and solvent molecules.  The method is thoroughly described in Refs.~\cite{Bohmer1990,Israels1994}; our implementation and additional terms that must be included in the free energy to account for dissociation of weak acid groups are described in Appendix B. We neglect spatial variations of segment densities in the directions lateral to the surface (and thus neglect the effect of ion-ion correlations), and determine the variation of densities in the direction normal to the surface.

   All calculations consider the conditions used for the experimental data shown in Fig.~\ref{fig:experimentalData}a, with pH=5, and 100 mM 1:1 salt with an additional 5mM 2:1 salt. The total surface density of functionalized molecules (neutral and charged) is $\sim 3$ nm$^{-2}$ at the nanoparticle surface\cite{Dragnea2008privcomm}.  Rather than explicitly modeling two species of surface molecules (neutral and acid-terminated), we vary the surface charge density $\sigc$ by changing the valency of the ionic end group.  To explore a wide range of surface charge we consider end groups with valencies $v\in [0,2]$; the experimental data in Fig.~\ref{fig:experimentalData} corresponds to the range $v\in[0,1]$. For $v=1$ the number of functionalized charge groups corresponds to $\sim 80\%$ of the charge on a BMV capsid. The lattice size and statistical segment length for polyelectrolyte and TEG molecules \cite{Kawakami2006,Oesterhelt1999,Kienberger2000} are set to $l=0.35$ nm (which roughly corresponds to the size of a single amino acid and the correlation length of water), and there are 0.5 charges per polyelectrolyte segment, which corresponds to the linear charge density  predicted by counterion condensation \cite{Manning1969,Muthukumar1996,Patra1999,Kundagrami2008}. Except for charge, all species are treated as chemically identical, with the Flory interaction parameter $\chi=0$ and dielectric permittivity $\epsilon=80$). Free energies are insensitive to the introduction of unfavorable interactions ($\chi>0$) between the alkane segments and hydrophilic species or species-dependent dielectric constants.  In the strong coupling limit,  the equilibrium density of positive charge due to adsorbed subunits  is not sensitive to statistical segment length, lattice size, polymer length or charge per segments.  The dissociation constant ($\pka$) for carboxylate groups is not known in the vicinity of the surface functionalized gold particle; we use $\pka=4.5$, so that approximately 25\% of the acid groups are dissociated for $\pH=5$ at full coverage (see below), consistent with Ref.~\cite{Gershevitz2004}. We do not explicity consider complexation of acid groups, a possibility suggested in that reference. We will also consider experiments in which the terminal group is a strong acid, with $\pka=0.0$.
\begin{figure} [bt]
\epsfig{file=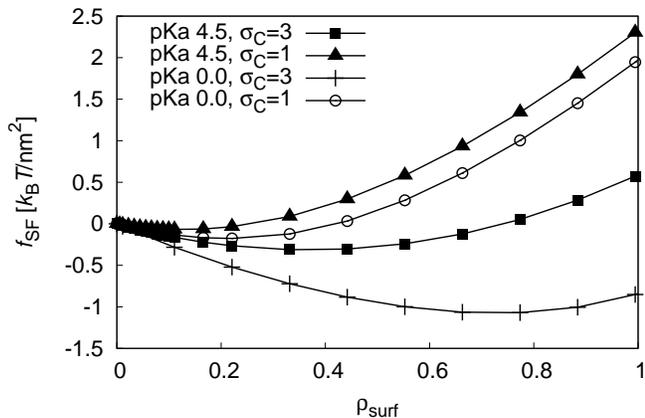,width=\textwidth/2}
\caption{\label{fig:Fexcess}
The free energy per area at the capsid surface for the adsorption of a concave polymer brush onto a spherical brush with opposite charge.  For varying surface densities of polyelectrolyte (normalized by the density in a complete capsid), $\rsurf$, the free energy relative to $\rsurf=0$ is shown for nanoparticle functionalization with weak acid groups ($\pka=4.5$) at $\sigc=3$ acid groups/nm$^2$ ($\blacksquare$), $\sigc=1$ ($\blacktriangle$) and strong acid groups at $\sigc=3$ (+), $\sigc=1$ (o).  There is 100 mM 1:1 and 5 mM 2:1 salt and $\pH=5$.}
\end{figure}

\begin{figure} [bt]
\epsfig{file=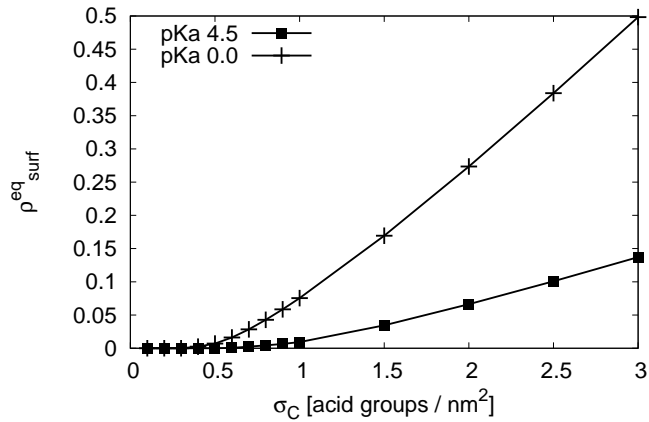,width=\textwidth/2}
\caption{\label{fig:Csurf}
The equilibrium surface density of adsorbed polyelectrolytes, $\csurf$, normalized by the surface density of polyelectrolyte on a core with a complete capsid. Predictions are shown for surface functionalization with weak acid ($\blacksquare$) and strong acid (+) groups. The bulk subunit concentration is 10 $\mu$M, with other parameters as in Fig.~\ref{fig:Fexcess}.}
\end{figure}

\emph{Free energies and equilibrium concentrations of adsorbed polyelectrolytes.}
  The equilibrium surface density of adsorbed polyelectrolyte can be calculated by considering a system for which there is free exchange between surface adsorbed polyelectrolytes and a bath containing polyelectrolyte, salt, and solvent.  The kinetic theory developed in Section~\ref{sec:kineticTheory}, however, requires the free energy for non-equilibrium adsorption densities. We therefore consider a restricted number of grafted molecules (which cannot exchange with the bath), as a function of surface charge and density of grafted molecules.    For each grafting density $\rsurf$, the spatial distributions of polyelectrolyte, TEG, ions, and solvent are determined by minimizing the total free energy.  The free energy, $\Fsf$, is calculated from  Eqs.~\ref{eq:freeEnergy} - \ref{eq:fexc}.  Note that the density of polyelectrolytes (Fig.~\ref{fig:layerHeights}) segments is nonmonotonic with distance from the capsid, consistent with studies of capsid assembly around nucleic acids \cite{Belyi2006,Siber2008}.

 Free energies for two functionalized charge densities with weak and strong acid groups are shown in  Fig.~\ref{fig:Fexcess} and  the equilibrium adsorption densities, $\csurf$, which correspond to the minimum excess free energy, are shown in Fig.~\ref{fig:Csurf}. The adsorption densities ($\rsurf$, $\csurf$) are normalized with respect to the density of dimer subunits in a complete capsid (0.09 subunits/nm$^2$ on the interior surface of the capsid with inner radius 8.9 nm).  The predicted adsorption densities for weak acids ($\pka=4.5$) are significantly lower than for the case of strong acids because the high local density of negative charges shifts the carboxylate dissociation equilibrium; for a carboxylate surface density  $\sigc=3/\text{nm}^2$ the average dissociated fraction shifts from 75\% for an isolated molecule to 35\%.  However, dissociation is enhanced as polyelectrolytes with positive charges adsorb and the average dissociation fraction on a core with a complete capsid is 78\%. As discussed in  Section~\ref{sec:ktResults}, this effect results in significantly different dependencies of assembly kinetics on $\csurf$ for weak and strong acids.

\begin{figure} [bt]
\epsfig{file=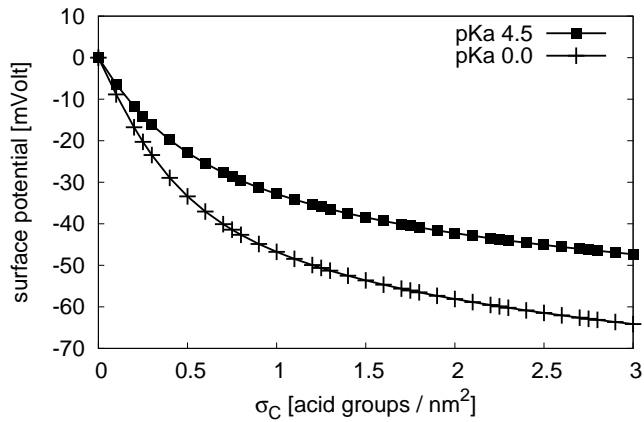,width=\textwidth/2}
\caption{\label{fig:surfacePot}
The electrostatic potential at the nanoparticle surface for varying surface functionalization ($\sigc$) with strong (+) and weak ($\blacksquare$) acid groups. There is 100 mM 1:1 and 5 mM 2:1 salt.}
\end{figure}

There are several reasons why our calculations may underestimate $\csurf$ at low $\sigc$ and for weak acid groups. Nanoparticle-protein interactions in addition to those involving the N-terminal arms could contribute to subunit adsorption. The calculated interactions involving N-terminal arms assume uniform charge densities in directions parallel to the surface, which will overestimate polyelectrolyte interactions at low surface densities, when the average distance between adsorbed subunits is greater than the polyelectrolyte radius of gyration; the approach described for neutral subunits in Appendix A might be appropriate in this regime. As discussed above, reducing restrictions on polyelectrolyte conformations due to the folded portion of capsid proteins results in somewhat higher adsorption densities for low surface charges. In addition, the $\pka$ of carboxylate groups and the effect of interactions between the charged species and the gold surface are poorly understood. For this reason, we examine assembly over a wide range of surface charges for both $\pka=4.5$ and $\pka=0$ to understand the interplay between surface charge, dissociation, and assembly. It would be desirable to experimentally measure the surface charge as a function of the fraction of charged TEG molecules on the surface; while it will be difficult to quantitatively verify the surface charge or corresponding magnitude of the electrostatic potential, the variation of the potential at the core surface with $\sigc$ (see  Fig.~\ref{fig:surfacePot}) is roughly consistent with preliminary experimental measurements of zeta potentials for functionalized nanoparticles \cite{Dragnea2008privcomm}.  While our calculations may underestimate $\csurf$, these approximations are less important at high coverages and hence should not significantly affect equilibrium packaging efficiency calculations.

\begin{figure} [bt]
\epsfig{file=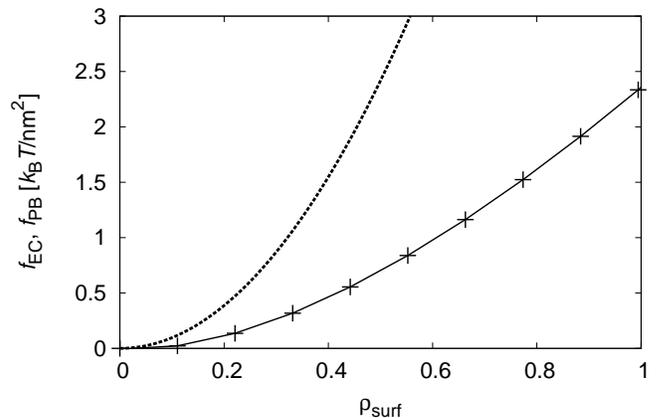,width=\textwidth/2}
\caption{\label{fig:SFempty}
The free energy per area at the capsid surface of an empty capsid due to electrostatic repulsions and ion and solvent entropy is shown as a function of the charge density on capsid subunits, as predicted by SF calculations (solid lines with symbols) and the linearized PB equation (dashed line). The calculations use 100 mM 1:1 salt and 5 mM 2:1 salt.  }
\end{figure}

\subsection{Empty-capsid free energies.}
\label{sec:SFemptyFreeEnergies}
We use the same approach, but without the core and TEG molecules, to estimate the electrostatic contribution to the free energy of forming an empty capsid. The free energy, $\FSFEC$, as a function of  polyelectrolyte density ($\rsurf$) is shown in Fig.~\ref{fig:SFempty}, where it is compared to the free energy calculated with the linearized Poisson Boltzmann equation for charge smeared on a spherical surface \cite{Kegel2004,Siber2007} with radius $\Rcap=8.9$ nm,
$f_\text{PB}=\lambda_{\text{B}}\lambda_{\text{D}} q_\text{s}^2 n_{\text{s}}^2 / [2 \Rcap( \lambda_{\text{D}}+\Rcap)]$,
with $n_\text{s}=90$ subunits, $q_\text{s}$=18 charges/subunit, $\lambda_{\text{B}}\approx 0.7$ nm the Bjerrum length of water, and $\lambda_{\text{D}}$ the Debye length.
Although the linearized Poisson Boltzmann equation was shown to agree closely with the full nonlinear calculation in Ref.~\cite{Siber2007} for this ionic strength, there is a large discrepancy between the PB and SF calculations; the SF calculations predict much lower free energies because the flexible tails adopt stretched conformations which reduce electrostatic repulsions (see also Ref.~\cite{Siber2008}).  We note, though, that we have not considered interactions between charges located in the structured region of capsid proteins.

Zlotnick and coworkers \cite{Ceres2002,Ceres2004,Zlotnick2007} have shown that capsid formation free energies can be fit using the law of mass action to experimental data for capsid and free subunit concentrations as the total capsid protein concentration is varied. The resulting free energies depend on $p$H as well as salt concentration \cite{Ceres2002} and likely include effects arising from hydrophobic as well as electrostatic interactions ~\cite{Kegel2004}.  We can subtract the polyelectrolyte free energy from the total capsid formation free energy derived from experimental data to estimate the free energy due to attractive subunit-subunit interactions, $\gh$, which may arise from hydrophobic contacts.  Following Ceres and coworkers \cite{Ceres2002}, we assume that each subunit in a capsid makes favorable contacts with $n^\text c=4$ neighboring subunits, so the attractive energy per subunit is
\begin{equation}
\gh = \gb n^\text{c}/2 - 4 \pi \Rcap^2 \FSFEC/N
\label{eq:definegh}
\end{equation}
For  $N=90$ protein dimer subunits in a $T$=3 BMV capsid with an inner capsid radius of $\Rcap=8.9$ nm \cite{Lucas2002},   we obtain $\FSFEC=2.35$ $\kt/\text{nm}^2$.  With a typical subunit-subunit contact free energy of $\gb=-5$ $\kt$ (-3 kcal/mol) \cite{Ceres2002,Ceres2004,Zlotnick2007}, this gives $\gh   \approx -36$ $\kt/\mbox{subunit}$ (compare to the estimate from experimental data for hepatitis B virus assembly by Kegel and van der Schoot \cite{Kegel2004} of an attractive interaction strength of $-13 \kt$ per subunit).

\section{Equilibrium theory for core-controlled assembly}
\label{sec:three}

 In this section we consider the thermodynamics for a system of subunits that can assemble into capsids around cores and assemble in bulk solution to form empty capsids.  The free energies due to electrostatic interactions for subunits adsorbed on surfaces or assembled into empty capsids developed in section~\ref{sec:two} are used to determine the relative free energies of core-associated capsid intermediates.

 \subsection{The thermodynamics of core-controlled assembly}
\label{sec:threep1}
We consider a dilute solution of capsid protein subunits with density $\CS$, and solid cores with density $\CC$.  Subunits can associate to form capsid intermediates in bulk solution or on core surfaces.  A complete capsid is comprised of $N$ subunits.  Zandi and van der Schoot \cite{Zandi2008} develop the free energy for core-controlled assembly in the context of capsids assembling around polymers, but neglect the contribution of partial capsid intermediates, which will be important for describing the kinetics of core-controlled assembly in the next section.  Here, we consider the free energy for a system of subunits, capsids, and partial capsid intermediates.  To simplify the calculation, we neglect the possibility of malformed capsids, but note that disordered configurations could be kinetically arrested or even the favored equilibrium state in the case of core-subunit interactions that are much stronger than the thermal energy $\kt$ \cite{Elrad2008,Nguyen2008b}.

The total free energy density is given by
\begin{equation}
F=\Fec + \Fcore
\label{eq:F}
\end{equation}
with $\Fec$ the free energy density of empty capsid intermediates
\begin{equation}
\Fec=\sum_{i=1}^N \rho_i \log(\rho_i a^3)-\rho_i + \rho_i G^\text{cap}_i
\label{eq:Fmt}
\end{equation}
where $\rho_i$ is the density of intermediates with $i$ subunits and $a^3$ is a standard state volume.  For simplicity we follow Zlotnick and coworkers \cite{Zlotnick1994,Endres2002}  and assume that there is only one intermediate for each size $i$, with a free energy $G^\text{cap}_i$ defined below in Eq.~\ref{eq:Gi}.

The free energy density of core-associated intermediates is given by
\begin{eqnarray}
\Fcore&=&\CC \bigg[ {\sum_{n=0}^N\sum_{m=0}^{N}} \sp{\prime}\Pnm\log(\Pnm \CC a^3) \nonumber \\
& & -\Pnm + \Pnm \Gcore_{n,m} \bigg]
\label{eq:Fc}
\end{eqnarray}
where $\Pnm$ is the fraction of cores with an adsorbed intermediate of size $m$ and with $n$ adsorbed but unassembled subunits.  The free energy for such a core is given by $\Gcore_{n,m}$ and is defined below. For simplicity, we assume that a core can have at most one assembled intermediate and a total of $N$ subunits can fit on a single core surface, where $N$ is the size of a complete capsid, so states with $m+n>N$ subunits have zero probability.  The prime on the second sum indicates that $m=0,2,3\dots N$ because, by definition, a state with $m=1$ has no assembled intermediates.

To obtain the equilibrium concentration of intermediates we minimize the total free energy subject to the constraints
\begin{equation}
\CC \sum_{n=1}^N{\sum_{m=0}^N}\sp{\prime}(n+m) \Pnm+\sum_{i=1}^N i \rho_i=\CS
\label{eq:CC}
\end{equation}
and
\begin{equation}
\sum_{n=1}^N{\sum_{m=0}^N}\sp{\prime} \Pnm=1
\label{eq:sumPnm}
\end{equation}
where Eq.~\ref{eq:CC} enforces that the total density of subunits is given by $\CS$.

Minimization yields
\begin{equation}
\rho_i a^3=\exp[-\beta(\Gcap_i - i \mu)]
\label{eq:rhoi}
\end{equation}
and
\begin{equation}
 \Pnm \CC a^3 =\exp[-\beta(\Gcore_{n,m}-(n+m)\mu - \mu_\text{C})]
\label{eq:Zcore}
\end{equation}
where $\beta=1/\kt$ is the inverse of the thermal energy, $\mu=k_{\text{B}}T\log(a^3 \rho_1)$  and $\mu_\text{C}=\kt \log(a^3 \CC P_{0,0})$ are chemical potentials for subunits and cores, respectively. 

\subsection{Free energies for empty-capsid assembly intermediates}
\label{sec:emptyFreeEnergy}

The free energy of an empty capsid intermediate with $i$ subunits is written as
\begin{equation}
\Gcap_i(\gb)=\sum_{j=1}^i (\ncontact_j \gb) - \kt \sdegen_i - i \kt \ssb
\label{eq:Gi}
\end{equation}
where $\ncontact_j$ is the number of new subunit-subunit contacts formed by the binding of subunit $j$ to the intermediate, $\gb$ is the contact free energy, $\sdegen$ accounts for degeneracy in the number of ways subunits can bind to or unbind from an intermediate (see the $s$ factors in Refs.~\cite{Zlotnick1994, Endres2002}), and $\ssb$ is a rotational binding entropy penalty.  For simplicity and to ease comparison with earlier works, we take $\ssb=0$ except when comparing theoretical predictions to simulation results (see Appendix A).  As discussed at the end of Section~\ref{sec:two}, the subunit-subunit contact free energy has been estimated from experiments for several viruses \cite{Ceres2002,Ceres2004,Zlotnick1999,Zlotnick2007} and depends on salt concentration and $p$H \cite{Ceres2002,Kegel2004}.

\subsection{Free energies for assembly intermediates on core surfaces}
\label{sec:coreFreeEnergy}


  The free energy for a core without a partial capsid is given by $\Gcore_{n,0}=\Ac \FSF(\sigc,\rsurf)$ with $\Ac$ the core area, $\sigc$ the functionalized acid group density, $\rsurf=n/N$ the normalized density of adsorbed subunits, and $\FSF$ calculated in  Section~\ref{sec:two}.  Likewise, the free energy for a complete capsid is $\Gcore=\Gcap_m + \Ac [\FSF(\sigc,1)-\FSFEC]$, where the polyelectrolyte contribution to the capsid free energy ($\FSFEC$) is subtracted because it is included in $\FSF$.

In the case of a core with a partially assembled capsid and adsorbed but unassembled subunits, there is an inhomogeneous distribution of charge on the surface, with higher charge density in the region of the assembled cluster.  We determine the free energy for such a core, with $n$ adsorbed subunits and an intermediate of size $m$, in the following schematic way. We assume the unassembled subunits spread on the surface of the core not occupied by the intermediate.  There are thus two `pools' of charge on the surface -- the partial capsid with a high subunit density $\rsurf=1$ and the remaining area of the core surface with a charge density spread uniformly over the area not covered by the intermediate , $A_m^\text{free}=\Ac(N-m)/N$.  The total electrostatic free energy is then given by

\begin{eqnarray}
\Gcs_{n,m} & = & \Gcap_m + \left(\Ac-A_m^\text{free}\right)\left[\FSF(\sigc,1) - \FSFEC\right]+ \nonumber \\
 & &  A_m^\text{free}  \FSF(\sigc,\frac{n \qs}{N-m})    \quad \mbox{for   } n+m\le N \nonumber \\
  &  & \infty  \qquad \qquad \qquad \qquad \mbox{for   } n+m>N
\label{eq:Gcselec}.
\end{eqnarray}
 In general there should be a term in $\Gcore$ to account for strain energy if the core curvature is not commensurate with that of the intermediate; we assume that capsid and core curvatures match.

\emph{Core encapsidation.}   Above a threshold or critical subunit concentration (CSC), the majority of subunits will be in capsids\cite{Zlotnick1994,Bruinsma2003,Hagan2006}.  Because core-subunit interactions can lead to high localized surface charge densities, the threshold concentration for assembly in the nanoparticle system can be below the CSC for spontaneous assembly.  Above the CSC, both empty and core-filled capsids can assemble.  Due to the entropy cost of assembling a capsid on a core, Eqs.~\ref{eq:Zcore} and \ref{eq:rhoi} show that there is a threshold surface free energy ($\Gcore_{N,0}-\Gcap_N \approx -k_{\text{B}}T \log (\CC a^3)$ for a stoichiometric amount of capsid protein and nanoparticles), below which empty capsids and free cores are favored over full capsids. Above the threshold surface free energy and above the CSC, nearly all cores will be encapsidated ($P_{0,N}\approx 1$) at equilibrium.   However, we will see that there can be a higher threshold surface free energy for efficient encapsidation kinetics.

\section{Kinetic theory for core-controlled assembly}
\label{sec:kineticTheory}

In this section we use the free energies developed in section \ref{sec:three} for capsids and capsid intermediates as the basis for a kinetic theory of core-controlled assembly.

Zlotnick and coworkers \cite{Zlotnick1994,Endres2002} have developed a system of rate equations that describe the time evolution of concentrations of empty capsid intermediates
\begin{eqnarray}
\frac{d \rho_1}{d t} &=& -2 \fsp s_2 \rho_1^2 -\sum_{i=2}^{N}\fsp s_{i+1} \rho_i \rho_1 + \bsp \rho_i \nonumber \\
\frac{d \rho_i}{d t}&=&\fsp s_i \rho_1 \rho_{i-1} - \fsp s_{i+1} \rho_1 \rho_i \qquad \qquad i=2\dots N \nonumber \\
& & -\bsp \rho_i + \bsp \rho_{i+1}
\label{eq:ktEmpty}
\end{eqnarray}
where $\fsp$ and $\bsp$ are respectively the forward and reverse binding rates and $s_i$ is a statistical factor that describes the number of different ways to transition from intermediate $i$ to $i+1$ \cite{Zlotnick1994,Endres2002}; $s_2$ corrects for double counting the number of pairwise combinations of free subunits.  Following Ref.~\cite{Endres2002}, we simplify the calculation by requiring that transitions between intermediates are only allowed through binding or unbinding of a single subunit and there is only one intermediate for each size $i$. We assume that the assembly rate constant $\fsp$ is independent of intermediate size $i$, but the number of contacts per subunit is not (see Table 1).

We extend this approach to describe simultaneous capsid assembly and adsorption to spherical cores, giving the following expression for the time evolution of core states

\begin{eqnarray}
\lefteqn{\frac{d P_{n,0}}{d t}=\rho_1 \Omega_{n-1,0} P_{n-1,0}+\om_{n+1,0}P_{n+1,0}}\nonumber \\
& & -\left(\rho_1 \Omega_{n,0}+\om_{n,0}\right)P_{n,0} \nonumber \\
& & + \w_{n-2,2}P_{n-2,2} + \Wspontbar_{2}P_{n,2} \nonumber \\
& &    - (W_{n,0} + \Wspont_{0})P_{n,0} \nonumber \\
\lefteqn{\frac{d P_{n,2}}{d t}=\Omega_{n-1,2}\rho_1 P_{n-1,2}+\om_{n+1,2}P_{n+1,2}} \nonumber \\
& & - \left(\rho_1 \Omega_{n,2}+\om_{n,2}\right)P_{n,2}  \nonumber \\
& & +W_{n+2,0}P_{n+2,0} +\w_{n-1,3}P_{n-1,3} \nonumber \\
& &+ W^{\text{spont}}_{0}P_{n,0} + \Wspontbar_{3}P_{n,3}  \nonumber \\
& &-\left(W_{n,2}+ \w_{n,2} + \Wspont_{2}+\Wspontbar_{2}\right)P_{n,2} \nonumber \\
 \lefteqn{\frac{d P_{n,m}}{d t} = \rho_1\Omega_{n-1,m} P_{n-1,m} \qquad \qquad \qquad m=3\dots N \nonumber} \\
& & +\om_{n+1,m}P_{n+1,m} -\left(\rho_1\Omega_{n,m}+\om_{n,m}\right)P_{n,m} \nonumber \\
&&+\ W_{n+1,m-1}P_{n+1,m-1}+\w_{n-1,m+1}P_{n-1,m+1} \nonumber \\
& & + \Wspont_{m-1}P_{n,m-1} + \Wspontbar_{m+1}P_{n,m+1}   \nonumber \\
& &-\left(W_{n,m}+ \w_{n,m} + \Wspont_{m} + \Wspontbar_{m}\right)P_{n,m}
\label{eq:ktCores}
\end{eqnarray}

The first two lines in each of Eqs.~\ref{eq:ktCores} describe adsorption and desorption from the core surface with respective rate constants $\Omega$ and $\om$, while the following lines describes binding and unbinding to a capsid intermediate.  Rate constants for binding and unbinding of adsorbed subunits to an adsorbed intermediate are denoted by $W$ and $\w$, respectively.  Finally, $\Wsp$ and $\Wspontbar$ are respectively the rate constants for the association (dissociation) of non-adsorbed subunits to (from) an adsorbed intermediate.  The latter process becomes important when an adsorbed capsid nears completion, and electrostatic repulsions render non-specific adsorption of subunits unfavorable, causing the surface capture and diffusion mechanism to be slow in comparison to the empty-capsid assembly mechanism.

Adsorption is calculated from $\Omega_{n,m}=k_{\text{ad}}\Phi_{n,m}$, with $k_\text{ad}$ the adsorption rate constant and $\Phi_{n,m}$ the probability that an adsorbing subunit is not blocked by previously adsorbed subunits.  For simplicity, we assume Langmuir kinetics, $\Phi_{n,m}=(N-m-n)/N$, but note that this is a poor approximation at high surface coverages (see Appendix A).

The desorbtion rate $\overline{\Omega}_{n,m}$ is related to the adsorption rate by detailed balance
\begin{equation}
\overline{\Omega}_{n,m}=\Omega_{n-1,m} \exp\left(\Gcore_{n,m}-\Gcore_{n-1,m}\right)/a^3
\label{eq:kad}.
\end{equation}

Assembly and disassembly of adsorbed subunits is described in a manner analogous to that used for empty capsids, with assembly rates given by
\begin{eqnarray}
W_{n,0}&=&\srf s_2 (n-1) \rho^\text{surf}_{n,0}  \\
W_{n,m}&=& \srf s_{n+1}\rho^\text{surf}_{n,m}D_{\text{C}} \qquad  m=2\ldots N-1 \nonumber
\label{eq:Wnm}
\end{eqnarray}
where the effective surface density is $\rho^\text{surf}_{n,m}=n/[a^3 (N-m)]$.

 Eq.~\ref{eq:Wnm}  states that subunit-subunit binding rates are proportional to the frequency of subunit-subunit collisions, which can be calculated from the Smoluchowski equation for the diffusion limited rate in three dimensions, with the density of subunits ($\rho^\text{surf}_{n,m}$) within a layer above the surface with thickness of the subunit size, $a$, or from the diffusion limited rate in two dimensions \cite{Berg1977} with a surface density given by $\rho_\text{2D}=\rho^\text{surf}_{n,m} a$.  The result of the two calculations differs only by a logarithmic factor that is of order 1 in this case (see appendix B of Ref.~\cite{Berg1977}).  Surface assembly rate constants ($\srf$) can be smaller than those for empty capsid assembly ($\fsp$), if interactions between subunits and the TEG layer impede diffusion or reorientation of adsorbed subunits. Association rates could also change in comparison to bulk solution if subunits take different conformations on the core surface; however, imaging experiments show that protein structures in nanoparticle-filled capsids are nearly identical to those in empty capsids \cite{Sun2007}, which suggests that subunits do not denature on the surface.

The disassembly rate is given by detailed balance
\begin{equation}
\overline{W}_{n,m}=W_{n+1,m-1} \exp\left(\Gcore_{n,m}-\Gcore_{n+1,m-1}\right)/a^3
\label{eq:WnmBar}.
\end{equation}

Similarly the association and dissociation rates for solubilized subunits are related by detailed balance:
\begin{eqnarray}
\Wsp_{0} &=& 2 \srf s_2 \rho_1 \nonumber \\
\Wsp_{m} &=& \srf s_{m+1} \qquad m=2\ldots N-1 \nonumber\\
\Wspontbar_m & = &\Wsp_m \exp \left(\Gcore_{n,m}-\Gcore_{n,m-1}\right)/a^3
\end{eqnarray}

The expression for the time evolution of empty capsid intermediates is now
\begin{eqnarray}
\lefteqn{\frac{d \rho_1}{d t} = -2 \fsp s_2 \rho_1^2 -\sum_{i=2}^{N}\fsp s_{i+1} \rho_i \rho_1
+ \bsp \rho_i +} \nonumber \\
& & \CC\sum_{n=1}^{N}\sum_{m=1}^{N}\left(\om_{n,m} - \Omega_{n,m} + \Wspontbar_{m} - \Wspont_{m}\right) P_{n,m} \nonumber \\
\lefteqn{\frac{d \rho_i}{d t}=\fsp s_i \rho_1 \rho_{i-1} - \fsp s_{i+1} \rho_1 \rho_i \qquad  \qquad i=2..N} \nonumber \\
& & -\bsp u_i + \bsp \rho_{i+1}
\label{eq:ktEmptyCores}.
\end{eqnarray}

We note that although Eqs.~\ref{eq:ktCores} and \ref{eq:ktEmptyCores} have the form of a Master equation, the finite concentration of subunits introduces a nonlinear dependence of the time evolution on state probabilities.

\section{Results}
\label{sec:ktResults}

\begin{figure} [bt]
\epsfig{file=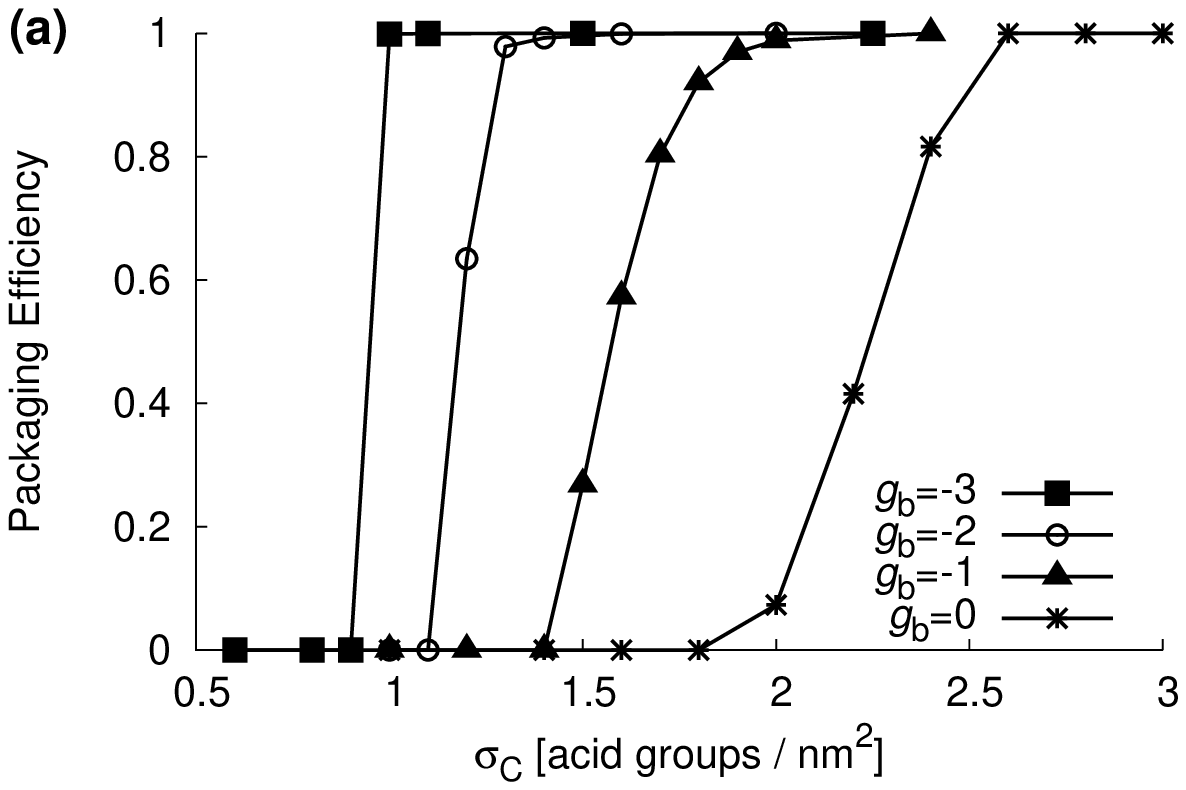, width=\textwidth/2}
\epsfig{file=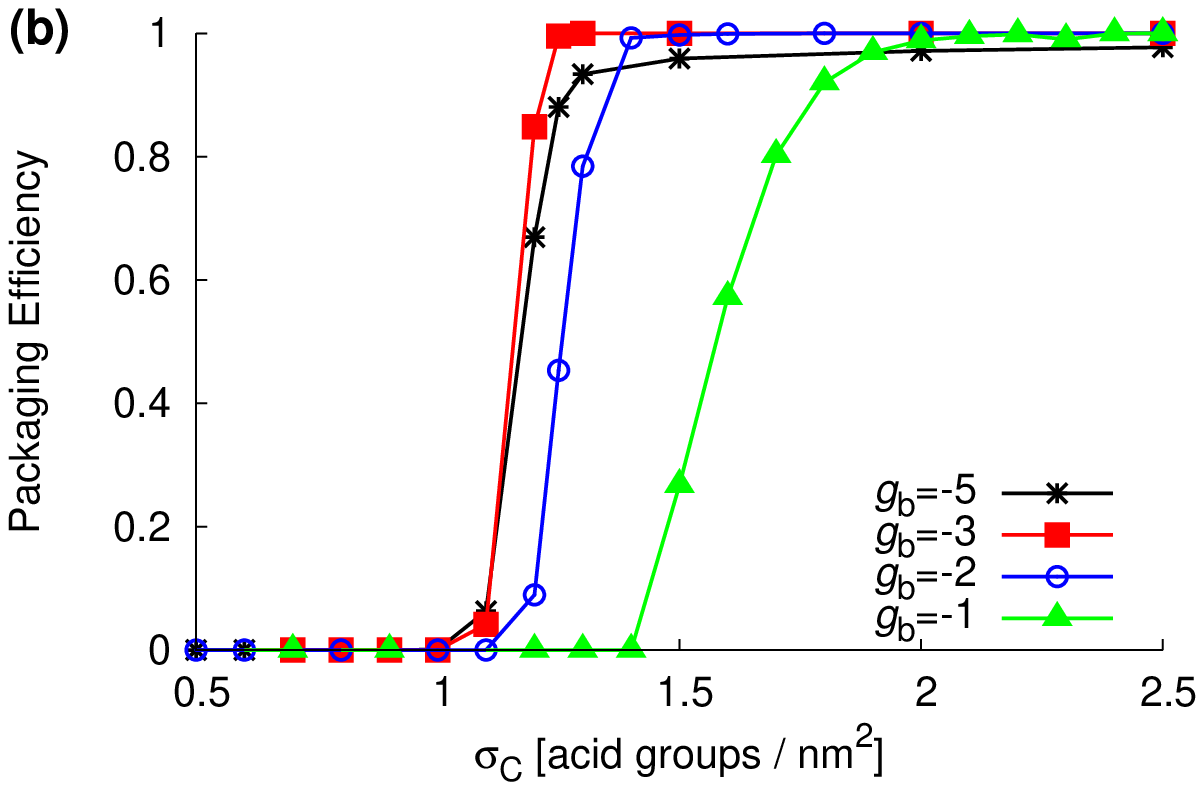, width=\textwidth/2}
\caption{\label{fig:peTheory}
 Packaging efficiencies, or the fraction of nanoparticles incorporated in well-formed capsids, at varying surface charge densities ($\sigc$) of weak acid groups predicted by {\bf (a)} the equilibrium theory and {\bf (b)} the kinetic theory at $10^5$ seconds.  Parameters are as listed in Fig.~\ref{fig:experimentalData} with $\kad=10^7 (\text{M}\cdot \text{s})^{-1}$, $\srf = 10^{4} \molsec$, and subunit binding free energies $\gb$ are indicated on the plots.  }
\end{figure}

 In this section we report the predictions of the equilibrium and kinetic theories for two experimentally observable quantities:  packaging efficiencies, or the fraction of nanoparticles encapsulated by well-formed capsids (estimated from TEM experiments), and time-resolved light scattering intensity. Preliminary experimental data for these quantities is shown in Fig.~\ref{fig:experimentalData}. Since there is not sufficient data to estimate the values of all unknown parameters, we make qualitative predictions about the effect of changing the experimental control parameters: the functionalized surface charge density $\sigc$, the stoichiometric ratio of capsid protein to nanoparticles $\npratio$, and the capsid protein concentration $\CCmolar$.  The important unknown parameters are the subunit-subunit binding energy, $\gb$, the dissociation constant of the functionalized charge groups, the rate constant for adsorption of subunits onto the nanoparticle surface, $\kad$, and the rate constant for the assembly of surface-adsorbed subunits, $\SRF$. The definitions and values of model parameters are given in Table~\ref{tab:parameters}.

\subsection{Packaging efficiencies}
\label{sec:PE}
We begin by discussing the relationship between the density of functionalized charge groups on nanoparticle surfaces, $\sigc$, and packaging efficiencies.  As discussed in Section~\ref{sec:intro}, the experimental measurements of packaging efficiencies were obtained from an experimental protocol in which the solution $\pH$ varied over time; because we do not know the dependence of the subunit-subunit binding energy $\gb$ on $\pH$, we only present predictions of the equilibrium and kinetic theories for packaging efficiencies at $\pH = 5$ for several binding energies.  We will compare these predictions to the currently existing data, but note that they should be compared to packaging efficiencies from experiments carried out entirely at $\pH = 5$, as was the case for the light scattering data shown in Fig.~\ref{fig:experimentalData}b.

As shown in Fig.~\ref{fig:peTheory}a, the equilibrium theory predicts effective packaging only above a threshold functionalized charge density, which depends on the binding free energy.  Effective packaging even occurs at a binding free energy $\gb=0$, showing that assembly on cores can occur under conditions for which spontaneous empty-capsid assembly is not favorable, as seen in experiments \cite{Sun2007} and simulations \cite{Hagan2008,Elrad2008}.  On core surfaces nanoparticle-subunit electrostatic interactions and attractive subunit-subunit interactions overcome subunit-subunit electrostatic repulsions to drive assembly ($\gh=-15.6$ kcal/mol per subunit  for $\gb=0$, Eq.~\ref{eq:definegh}).

 	As shown in Fig.~\ref{fig:peTheory}b, for weak subunit-subunit interactions ($\gb\ge -2$ kcal/mol) the kinetic theory predictions at $10^5$ seconds are essentially identical to the equilibrium results. As subunit-subunit interactions increase in magnitude, however, the predicted threshold charge density becomes larger than the equilibrium value, and even begins to increase for $\gb<-3$.  With stronger subunit-subunit binding interactions, the assembly of empty capsids competes with assembly on core surfaces by depleting the concentration of free subunits.  Assembly on cores remains favorable at high surface charge densities because initial subunit adsorption leads to rapid nucleation, and subsequent nonspecific adsorption enhances capsid growth rates (see below).  The predicted packaging efficiencies at long times are relatively insensitive to variations of the kinetic parameters $\kad$ and $\SRF$.

Interestingly, the threshold surface charge density predicted by the kinetic theory, $\sigc\approx 1.2$, is nearly independent of the subunit binding energy for $\gb \le 2$ kcal/mol, and comparable to the lowest experimental charge densities with measurable packaging efficiencies (Fig.~\ref{fig:experimentalData}). However, the experimental results show a gradual increase in packaging efficiency as the functionalized charge density increases.  We do not observe such a gradual increase in packaging efficiencies except for shorter observation times or an extremely low surface assembly rate constant.  This discrepancy could arise because we have assumed monodisperse cores that are perfectly matched to the capsid size, or it may indicate the presence of kinetically frustrated states (see below).

\begin{figure} [bt!]
\epsfig{file=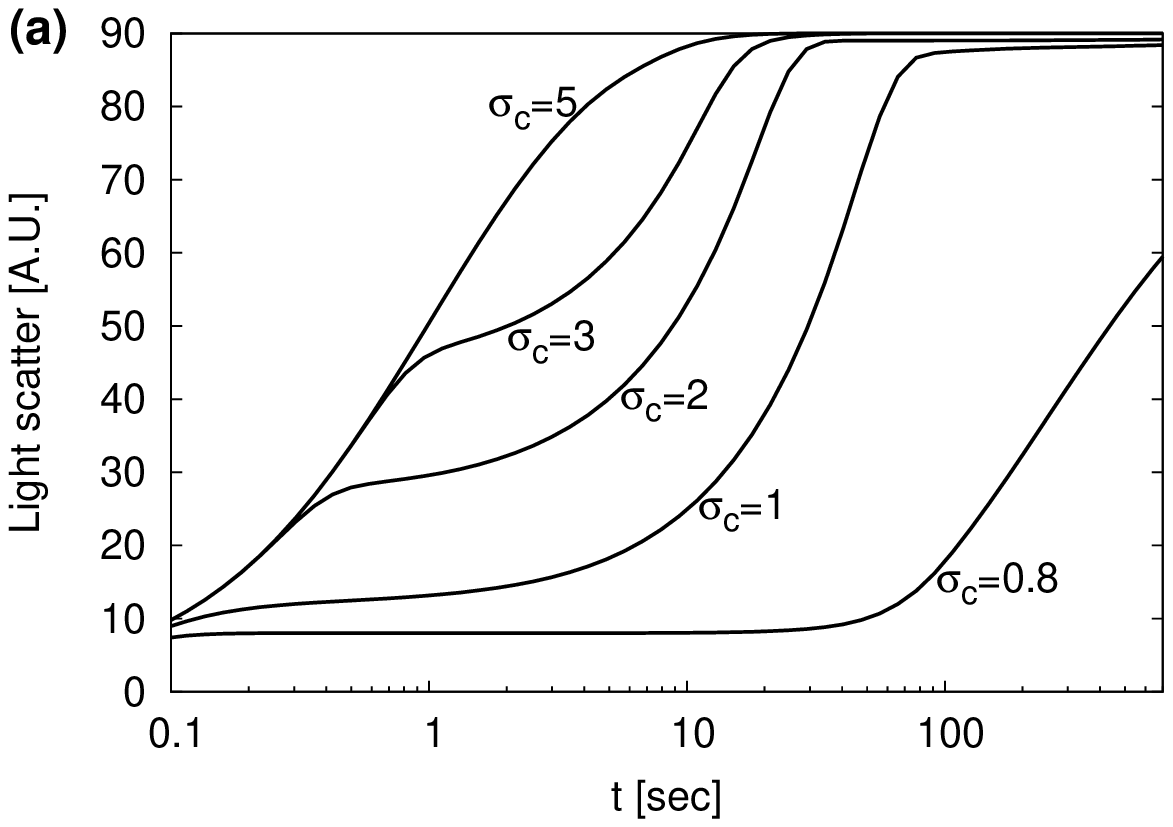,width = 0.48 \textwidth}
\epsfig{file=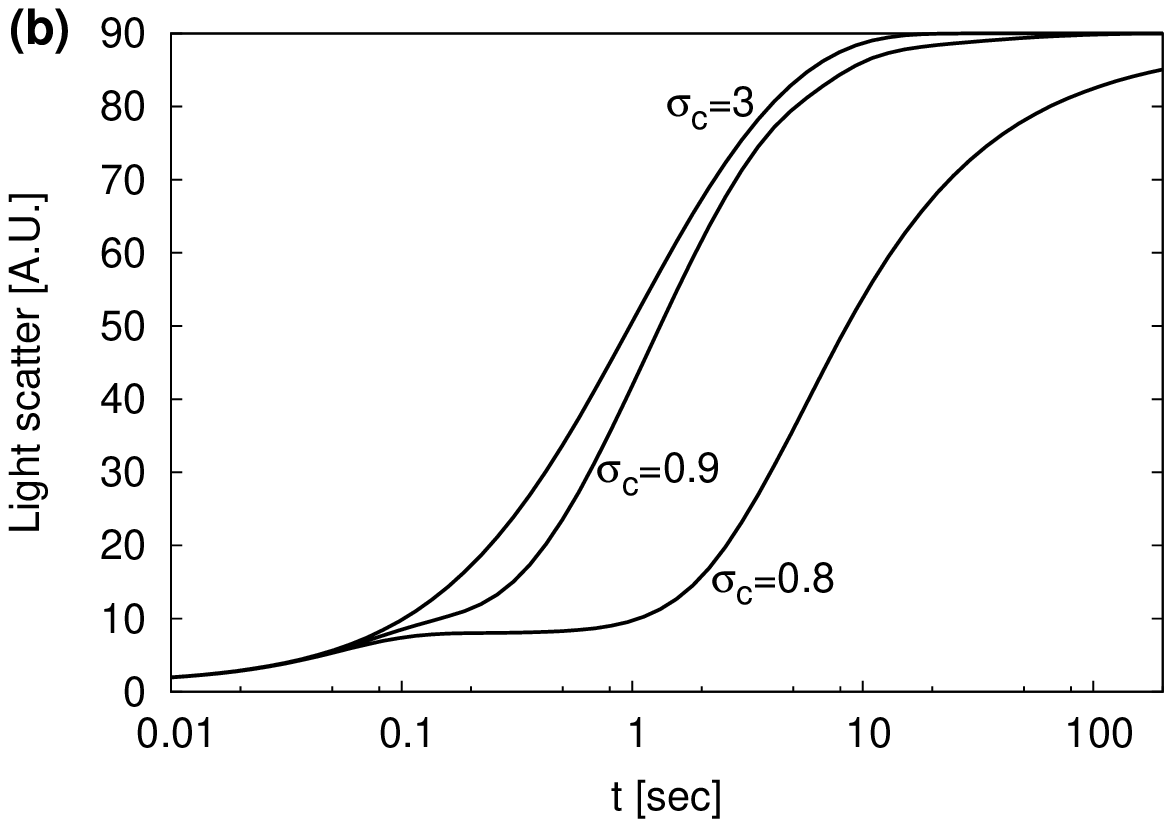,width = .48 \textwidth}
\epsfig{file=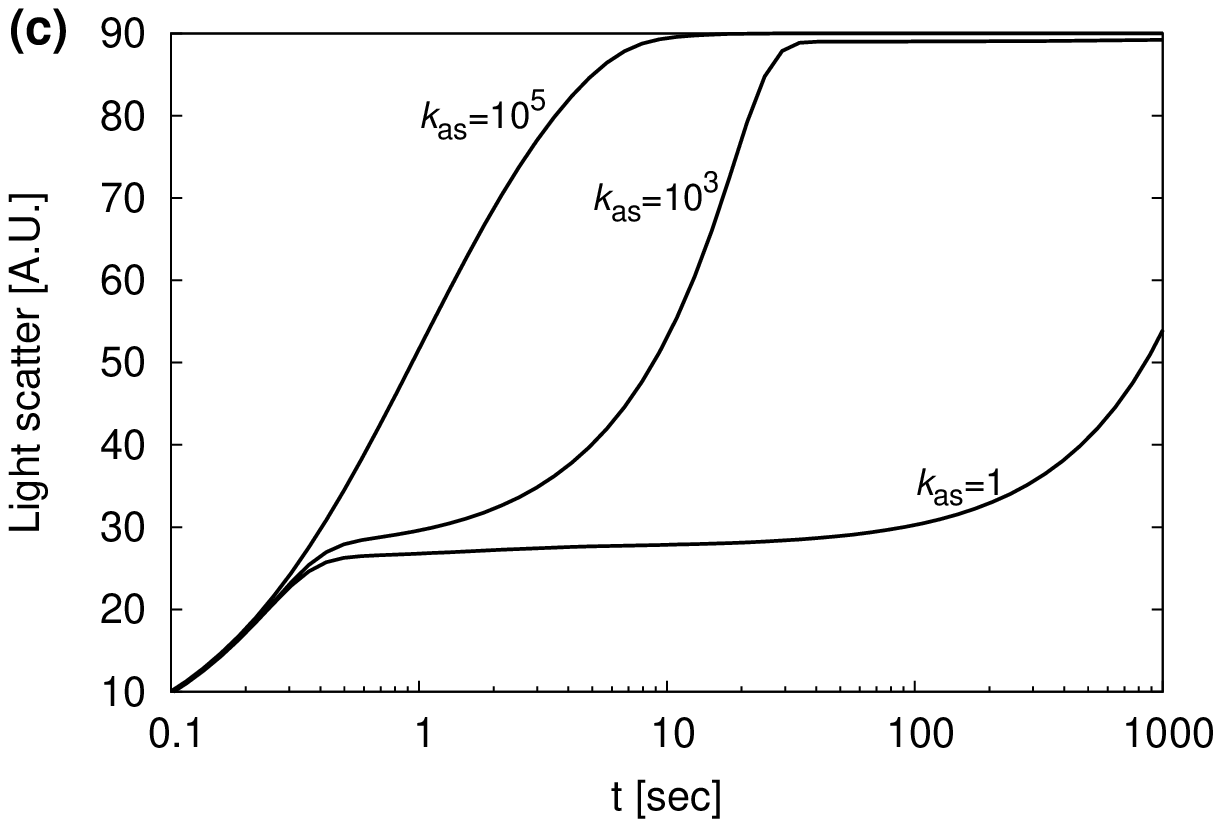,width = .48 \textwidth}
\caption{\label{fig:lightScatter}
The predicted time dependence of light scattering for functionalization with strong acid groups.  {\bf (a)} and {\bf (b)} Indicated functionalization densities with rate constants {\bf (a)} $\srf=10^{2} \molsec$ and {\bf (b)} $\srf=10^4 \molsec$.  {\bf (c)} Indicated surface assembly rate constants,  with $\sigc=3$ nm$^{-2}$.  Other parameters for {\bf (a)}-{\bf(c)} are $\kad=10^7 (M\,s)^{-1}$,$\npratio=1.33$, $\CB=10 \mu$M, and $\gb=-2$ kcal/mol.}
\end{figure}

\begin{figure} [bt!]
\epsfig{file=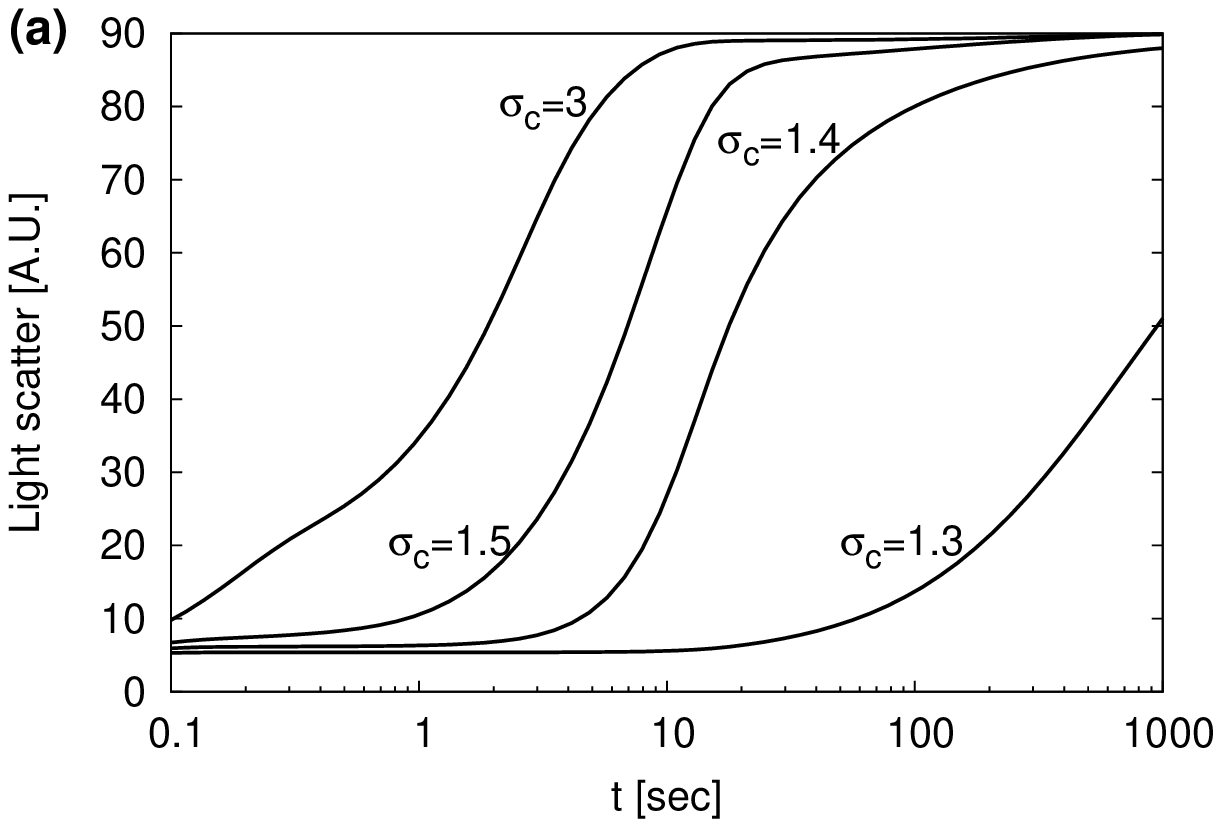,width = .48 \textwidth}
\epsfig{file=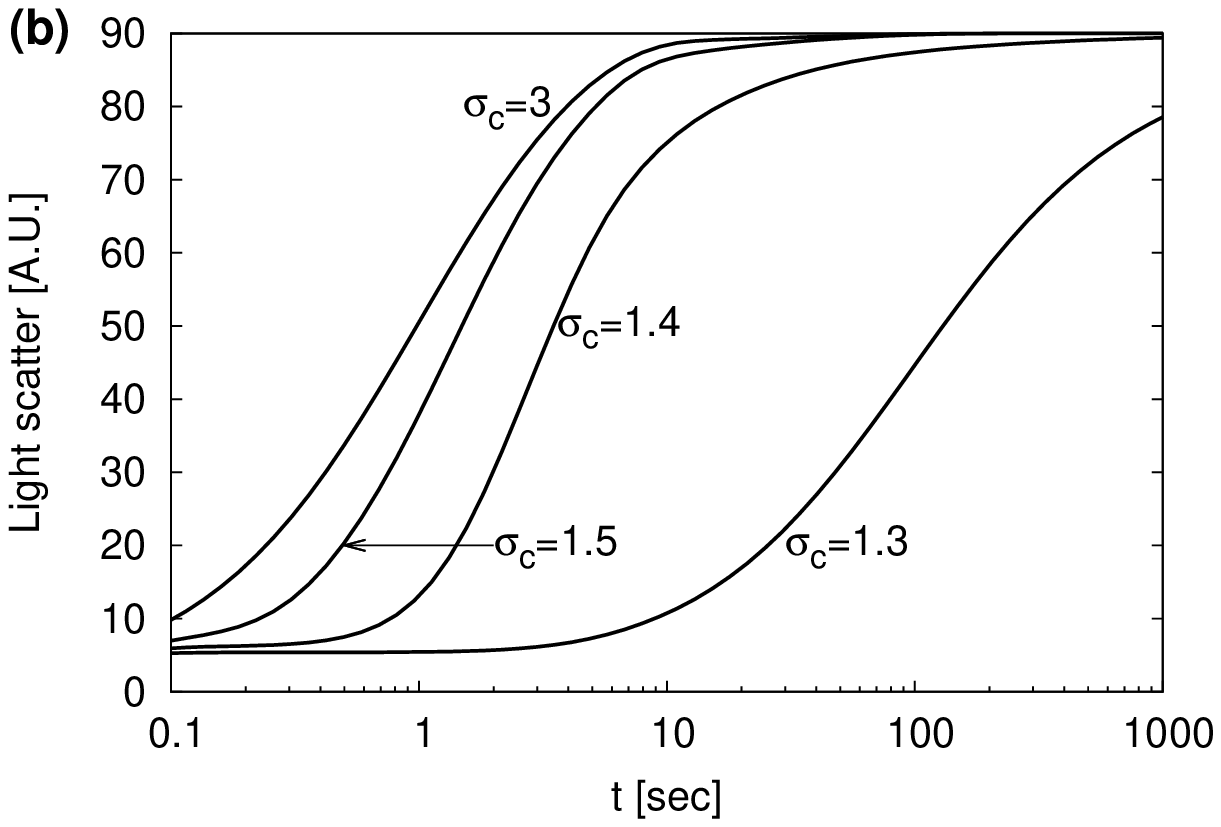,width = .48 \textwidth}
\epsfig{file=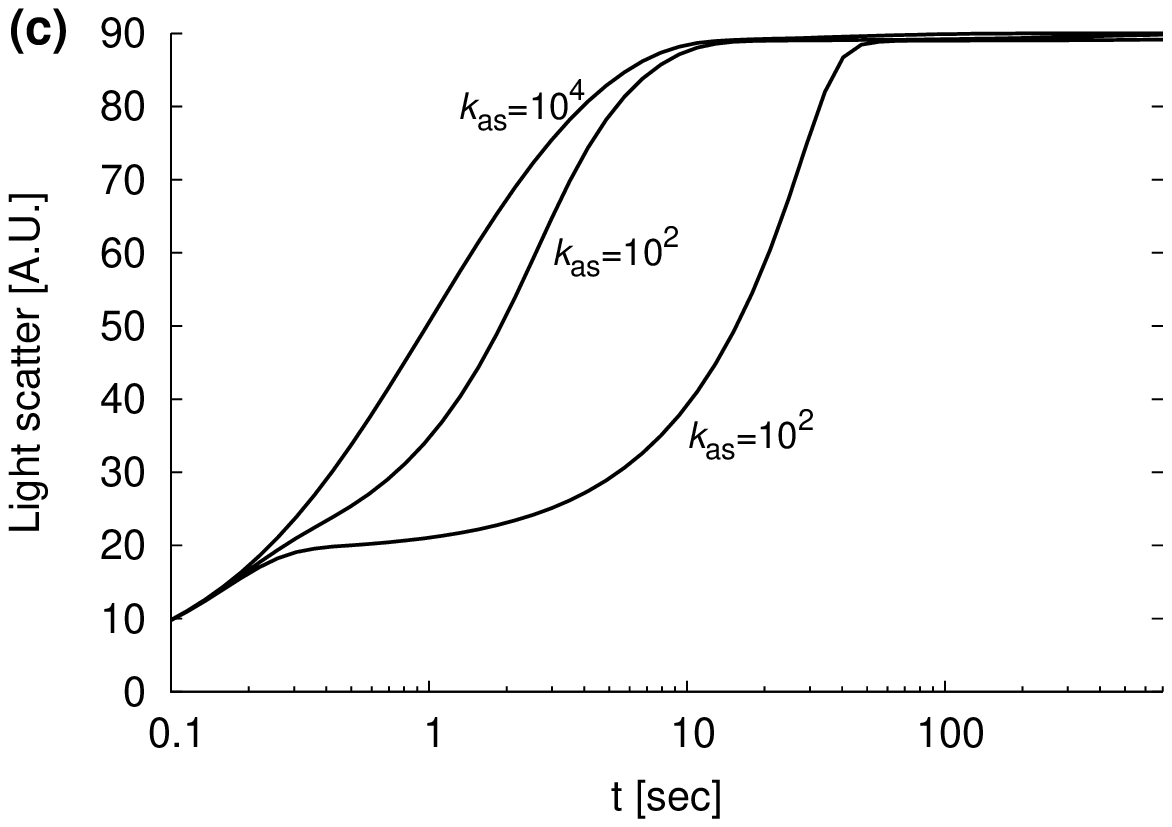,width = .48 \textwidth}
\caption{\label{fig:lightScatterPka}
The predicted time dependence of light scattering for surface charge groups with $\pka=4.5$ and $pH=5$, for {\bf (a)} indicated functionalization densities $\srf=10^{3} \molsec$, {\bf (b)} indicated functionalization densities with $\srf=10^4 \molsec$, and {\bf (c)} indicated surface assembly rate constants with $\sigc=3$ nm$^{-2}$.  Other parameters are as in  Fig.~\ref{fig:lightScatter}.}
\end{figure}

\subsection{Assembly kinetics on core surfaces}
In this section we explore the effect cargo-capsid protein interaction strength on assembly kinetics.  As discussed in Section~\ref{sec:intro}, assembly kinetics can be monitored in experiments with time-resolved dynamic light scattering; preliminary experimental data for the highest functionalized charge density is shown in Fig.~\ref{fig:experimentalData}.  Because it is unlikely that the light scattering signal can distinguish between disordered and assembled subunits on core surfaces, we estimate light scattering from the kinetic theory by calculating the mass-averaged amount of adsorbed and assembled ($n+m$) subunits on core surfaces.  We simplify the presentation by varying only the charge density ($\sigc$) and the surface assembly rate constant ($\srf$), with physically reasonable values for the subunit binding free energy, $\gb=-2$ kcal/mol, and the the subunit adsorption rate constant, $\kad=10^7 (\text{M} \cdot \text{s})^{-1}$ \cite{kadNote}.  As discussed in section Section~\ref{sec:kineticTheory},  the effect of reduction in dimensionality due to adsorption onto a surface is accounted for in the kinetic theory, but assembly rate constants could still be dramatically different from those in bulk because of impeded diffusion of adsorbed subunits.  We therefore explore a wide range of surface assembly rate constants $\SRF$.

Predicted light scatter as a function of time is shown for various functionalized charge densities $\sigc$ and surface assembly rate constants $\srf$ in Figs.~\ref{fig:lightScatter} and ~\ref{fig:lightScatterPka} for functionalization with strong and weak acids\cite{strongAcidNote}, respectively. The analysis is easiest for the case presented in Fig.~\ref{fig:lightScatter}a  ($\srf=10^2 \mbox{M s}^{-1}$), for which assembly on the core surface is slow compared to subunit adsorption. For each value of $\sigc$, there is a rapid initial rise in light scatter, followed by a charge-dependent plateau, and finally a slow increase to saturation. The initial fast kinetics correspond to nonspecific subunit adsorption (i.e. with little or no assembly) and the plateau occurs when adsorption saturates at what would be the equilibrium surface density if there were no assembly ($\csurf$, see Section~\ref{sec:two}).  If adsorbed subunits assemble into a relatively stable intermediate, the chemical potential of remaining adsorbed subunits decreases, resulting in further adsorption and increased light scatter.  This process continues as additional subunits bind to the growing intermediate, until assembly stops.

At high functionalization densities, nonspecific subunit adsorption approaches the density of a complete capsid; the high local or surface concentration of subunits enables rapid nucleation and assembly and thus the light scatter reaches its final saturation value at relatively short times.  At the lower functionalization densities, nonspecific adsorption saturates at smaller surface concentrations, with corresponding smaller initial plateau heights for the light scatter.  Nucleation times increase dramatically as the surface concentration is reduced (see the discussion in Ref.~\cite{Hagan2008}), resulting in the long plateau before assembly completes.

For a larger assembly rate constant $\srf=10^4 \mbox{M s}^{-1}$, the nonspecific adsorption and assembly phases can only be clearly distinguished at low functionalized charge densities, as shown in Fig.~\ref{fig:lightScatter}b; at larger $\sigc$ adsorption and assembly are concomitant.  For weak acid groups (Fig.~\ref{fig:lightScatterPka}), equilibrium surface densities increase only slowly with $\sigc$ ($\csurf\le0.15$, see Fig.~\ref{fig:Csurf}) because the local concentration of negative charge disfavors carboxylate dissociation.  There is still a strong dependence of overall assembly rates on $\sigc$, however, because the positive charge from capsid proteins drives additional carboxylate dissociation (see Section~\ref{sec:two}). Thus, assembly effectiveness cannot be predicted from $\csurf$ alone, as was found for simulations of neutral subunits in Ref.~\cite{Elrad2008}.

\emph{Cores can enhance assembly rates.} From the preceding discussion, it is clear that functionalized charge leads to an increased concentration of capsid proteins nanoparticle surfaces relative to that in bulk solution, which can increase capsid nucleation rates and even induce assembly below the threshold protein concentration for formation of empty capsids. Because nonspecific adsorption onto core surfaces can be much faster than subunit-subunit binding rates, the core surface can also enhance assembly rates during the growth phase by increasing the effective subunit-subunit interaction cross-section. This effect was hypothesized for RNA by Hu et al.~\cite{Hu2007b} and is reported for simulation results in Ref.~\cite{Hagan2008}.

\emph{Assembly rates decrease as capsids near completion.}  Net assembly rates decrease as capsids near completion because adsorption of subunits onto cores with large partial capsids is impeded by two effects. First, the excluded volume of the partial capsid decreases adsorption rates; a similar effect has been observed in simulations representing empty capsid assembly, but the magnitude of rate reduction seems to be model dependent (see Refs.~\cite{Nguyen2007,Hagan2008}).  Electrostatic repulsions with adsorbed and/or assembled subunits on the core surface can have a larger effect on subunit adsorption. When the subunit surface density is large compared to $\csurf$, the surface capture and diffusion assembly mechanism that enables rapid core-controlled assembly becomes inefficient. Even though direct association of free subunits onto core-bound intermediates is included in the kinetic theory, there is a reduction in assembly rates as capsids near completion; this effect is most noticeable for surface functionalization with weak acid groups (Fig.~\ref{fig:lightScatterPka}a), or low fractions of strong acid groups ($\sigc < 2$ nm$^{-2}$).  The decreased assembly rates are reflected in slow growth of light scatter as assembly nears completion.

\begin{figure} [bt]
\epsfig{file=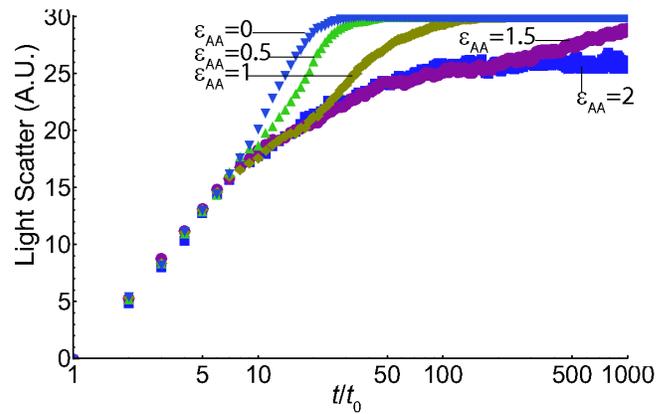,width = .48\textwidth}
\caption{\label{fig:adaptiveAssemblyLS}
Estimated light scatter from simulations in which subunits assemble around model nanoparticle with a size mismatch.  The preferred empty-capsid morphology is a T=3 capsid, but the lowest free energy configuration is a T=1 capsid assembled around the T=1-size nanoparticle. Estimated light scatter is shown as a function of time for different values of the parameter $\epsilon_\text{AA}$ (in units of $\kt$), which gives the free energy cost for adopting a subunit conformation consistent with a T=1 morphology. As described in Ref.~\cite{Elrad2008}, the larger values of $\epsilon_\text{AA}$ lead to the formation of frustrated nanoparticle-associated partial capsids with a T=3 morphology, which cannot close around the nanoparticle and block assembly into the lowest free energy configuration.  The data is replotted from simulations reported in Fig. 6 of Ref.~\cite{Elrad2008}, with the nanoparticle-subunit interaction energy $\epsilon_\text{s}=8 \kt$.
}
\end{figure}

\emph{Comparison with experimental data.} As discussed above, there is not sufficient experimental data to estimate values for the unknown parameters; however, the predicted light scatter plots share some features with the experimental light scatter data shown in Fig.~\ref{fig:experimentalData}b. In particular, for some parameter sets light scatter increases rapidly at short times, without the lag time seen for empty capsid assembly \cite{Zlotnick1999,Casini2004,Chen2008}. It is tempting to attribute the initial rise to nonspecific adsorption that does not saturate until nearly complete coverage, as seen for strong acid surface groups (Fig.~\ref{fig:lightScatter}a), but the calculation with weak acid groups predicts values of $\csurf$ that are well below complete coverage even for $\sigc=3$ nm$^{-2}$.  Interestingly, the predicted light scatter curves in this case with fast assembly kinetics ($\srf=10^4 \molsec$, Fig.~\ref{fig:lightScatter}a) still demonstrate a rapid initial rise followed by slow saturation; the initial rise corresponds to rapid nucleation and assembly concomitant with adsorption.  Thus, as discussed in the next section, comparison of theoretical predictions and experimental light scatter data at varying functionalized charge density will be necessary to unequivocally determine the origin of the initial light scatter phase.

The experimental light scattering data (Fig.~\ref{fig:experimentalData}b) appears to demonstrate a logarithmic increase in light scattering between 20 and 200 seconds. While the theoretical light scatter curves also demonstrate slow growth near completion (see above) due to subunit-subunit electrostatic repulsions, this effect is limited to the last five or 10 subunits.  We did not find any parameter sets for which predicted light scatter grew logarithmically over several decades in time, although our parameter search was not exhaustive. Intriguingly, the predicted light scatter from simulations in which frustrated off-pathway intermediates block assembly \cite{Elrad2008} do show logarithmic growth over several decades (see  Fig.~\ref{fig:adaptiveAssemblyLS}).

\subsection{Concentration effects}
\begin{figure} [bt]
\epsfig{file=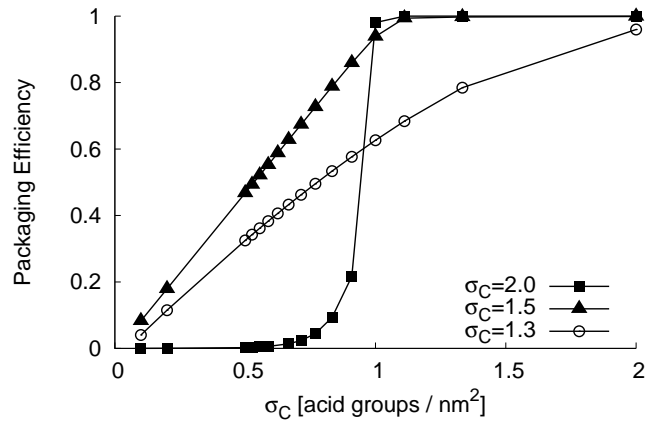,width = \textwidth/2}
\caption{\label{fig:CoreRatio}
Packaging efficiencies depend on the capsid protein/core stoichiometric ratio.  The predictions of the kinetic theory for the variation of packaging efficiencies  with the capsid protein/core stoichiometric ratio, $\npratio=\CB/N \CCmolar$ at a fixed subunit concentration of $\CB=10 \mu$M, are shown for functionalization with weak acid groups,  with other parameters as in Fig.~\ref{fig:peTheory}.
}
\end{figure}

\emph{Varying nanoparticle concentrations.} We first examine assembly effectiveness as the nanoparticle concentration is varied at fixed subunit concentrations.   We define the capsid protein-nanoparticle stoichiometric ratio as $\npratio=\CB/(N \CCmolar)$ so that $\npratio=1$ when there is exactly enough capsid protein to assemble a complete capsid on every core.  The variation of packaging efficiencies with the stoichiometric ratio is shown in Fig.~\ref{fig:CoreRatio}; the predicted curves are sigmoidal in shape, much like the experimentally observed packaging efficiencies shown in Fig. 1A of Ref.~\cite{Sun2007}.  Packaging  efficiencies for $\npratio \lesssim 1$ are typically lower than the equilibrium values obtained by solving Eqs.~\ref{eq:rhoi} and \ref{eq:Zcore},  because initially  capsid proteins undergo nonspecific adsorption onto every core, which depletes the concentration of free subunits.  The formation of complete capsids then requires desorption and subsequent re-adsorption onto cores with large intermediates.  The timescale for subunit desorption increases with functionalized surface charge density and therefore the kinetic packaging efficiency is nonmonotonic with respect to $\sigc$ for $\npratio \leq 1$.  This scenario is consistent with the slow assembly kinetics experimentally observed for capsid assembly in the presence of RNA \cite{Johnson2004}.

\begin{figure} [bt]
\epsfig{file=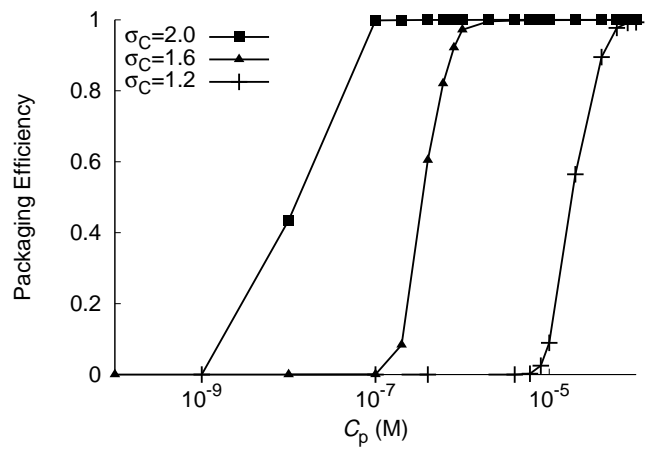,width = \textwidth/2}
\caption{\label{fig:u0}
Packaging efficiencies depend only weakly on the initial subunit concentration, $\CB$, for fixed $\npratio$.  The predictions of the kinetic theory for the variation of packaging efficiencies  with subunit concentration $\CB$ are shown for functionalization with weak acid groups,  with other parameters as in Fig.~\ref{fig:CoreRatio}.
}
\end{figure}

\emph{Varying subunit concentrations.} As shown in Fig.~\ref{fig:u0}, at $\sigc\ge2$ acid groups/nm$^{2}$ the kinetic theory predicts a much weaker dependence of assembly effectiveness on subunit concentration than has been seen for empty-capsid assembly.  This  trend arises because the chemical potential of adsorbed subunits is dominated by electrostatics rather than subunit translational entropy.

\section{Discussion}
\label{sec:discussion}
{\bf The feasibility of estimating the unknown parameters.}  The results in the last section are consistent with some features of the preliminary experimental data,  and the predicted dependencies of packaging efficiency and light scattering on the functionalized charge density and nanoparticle-protein stoichiometric ratio can be qualitatively compared with additional experimental data. Quantitative comparisons of experimental data, however, will require estimates for at least three currently unknown parameters, the subunit-subunit binding energy $\gb$, the assembly rate constant on core surfaces $\srf$, the subunit adsorption rate $\kad$. In Appendix A, we show that, with good estimates for the unknown parameters, the theoretical predictions agree reasonably with simulation results for a model of core-controlled assembly.  While it is not certain that a simplified theory with a few parameters will exactly match experimental data from a system with thousands of distinct intermediates, each of which could have different rate constants, rate equations with a similar level of coarse graining have successfully explained many features of empty-capsid assembly (e.g.~\cite{Zlotnick2007,Zlotnick1994,Endres2002,Zlotnick2000,Ceres2002,Singh2003}). Given the number of experimental control parameters in the nanoparticle-capsid assembly system, it should be possible to generate sufficient experimental data to further test the qualitative trends predicted here, and to make quantitative estimates for the unknown parameters through data fitting.  With estimated values for $\srf$ and $\kad$, it would be possible to predict the assembly state of subunits on core surfaces during the initial rise in light scatter, which would be challenging to determine with experiments alone.

In addition to measuring packaging efficiencies and light scattering data  over a range of functionalized charge densities (with both weak and strong acid groups) and capsid protein-nanoparticle stoichiometric ratios, the theoretical predictions in the last section suggest additional experiments that could be useful to provide independent estimates of some unknown parameters. In particular, the predicted light scatter curves in Figs.~\ref{fig:lightScatter} and \ref{fig:lightScatterPka} show that adsorption and assembly happen simultaneously for many parameter sets. Measuring light scatter of assembly incompetent proteins in the presence of functionalized nanoparticles would enable independent estimation of the subunit adsorption rate and the equilibrium subunit surface charge density $\csurf$ ( Fig.~\ref{fig:Csurf}).  In addition to varying the functionalized charge density, the subunit-nanoparticle interaction could modified by mutations to the N-terminal arm on capsid proteins\cite{Aniagyei2009}; Tang et al. \cite{Tang2006} have shown that cowpea chlorotic mottle virus capsid proteins with 34 residues deleted from the N-terminal arms are assembly competent (although they lose selectivity for the native capsid structure).

\emph{Identifying metastable disordered states.}
We make several important approximations in this work.   Most significantly, the theory considers only only intermediates consistent with the native capsid geometry, while experiments \cite{Dragnea2008privcomm} and simulations \cite{Elrad2008,Nguyen2008b} indicate that other capsid morphologies and/or asymmetric, malformed structures can be kinetically or even thermodynamically favored when core-subunit interaction strengths are large compared to the thermal energy $\kt$.  These effects could be accounted for in the theory by extending the state space to include structures other than complete and partial well-formed capsids, and by introducing subunit diffusion constants that depend on the strength of core-subunit interactions. If the present theory proves capable of describing experiments with parameters for which predominately well-formed capsids assemble, however, inconsistencies between theoretical predictions and experimental measurements for other system parameter values could identify kinetic traps.  For example, the calculated packaging efficiencies show a less gradual increase with functionalized charge density than measured in experiments and predicted light scatter does not fully capture the a logarithmic increase in the experimental light scatter curve.  Both of these discrepancies could be explained by frustrated off-pathway of assembly intermediates or slow subunit diffusion rates, as evidenced by predicted light scatter from simulations which do demonstrate frustrated assembly (Fig.~\ref{fig:adaptiveAssemblyLS}).  Confirming this possibility will require additional experimental data.

\begin{table*}[hbt] 
{\footnotesize
\begin{tabular}{|l|l|l|}\hline
Parameter & Value & Definition  \\ \hline
$N$ & 90 & Number of subunits in a complete $T$=3 capsid \\
$\Ac$& 995.4 nm$^2$ & Inner surface area of a complete capsid (Section~\ref{sec:coreFreeEnergy}\\
$a$ & 4.2 nm & Subunit diameter \\
$\gb$ & $[-5,0]$ kcal/mol & Binding free energy per subunit-subunit contact  (Eq.~\ref{eq:Gi}) \\
$\gh$ & $[-10.6,-5.6]$ kcal/mol &Free energy per subunit-subunit contact due to attractive interactions (Eq.~\ref{eq:definegh}) \\
$k_\text{ad}$ & $10^7$ $(M s)^{-1}$ &   Subunit adsorption rate constant \\
$\fsp$  & $10^5 \molsec$ & Empty capsid assembly rate constant \cite{Johnson2005} (Eq.~\ref{eq:ktEmpty})\\
$\srf$ & $[10^{2},10^5] \molsec$ &  Surface assembly rate constant (Eq.~\ref{eq:Wnm}) \\
$\sigc$ & $[-3,0]$ charge/(nm$^2$ gold surface)  & Surface-density of functionalized charge  \\
$\qs$& 18 &  Number of positive charges per protein-dimer subunit \\
$\CB$ & 10 $\mu$M & Total subunit concentration \\
$\npratio$ & $[0.1,10]$ & Capsid protein--nanoparticle stoichiometric ratio, $\npratio=\CB/(N \CCmolar)$ \\
$\CCmolar$  & $[0.11,1.1]$ $\mu$M & Nanoparticle concentration\\
\hline
\end{tabular}
}
\caption{\label{tab:parameters}
Parameter values used for  calculations in this work.
The parameters used for capsid assembly are as follows.  The number of subunit-subunit contacts are $\ncontact_j=1$ for $j\le\nnuc$, $\ncontact_j=2$ for $j\in (\nnuc,N-\nnuc+1]$, $\ncontact_{N-1}=3$ for $j\in (N-\nnuc,N)$, and $\ncontact_N=4$. This choice obviates the need for different nucleation and elongation rate constants \cite{Endres2002}. The nucleus size is $\nnuc=5$.  The forward and reverse reaction degeneracies are $s_2=\sbar_2=2$, $s_j=\sbar_j=3$ for $3\le j < N$ (the average value calculated from simulations in Ref.~\cite{Hagan2008}) and $s_N=1$, $\sbar_N=N$.  The binding degeneracy entropy  (Eq.~\ref{eq:Gi})  is $\sdegen=k_\text{B} \log (s_j/\sbar_j)$.\\
 }
\end{table*}

\section{Conclusions}
\label{sec:conclusions}
In summary, we develop simplified thermodynamic and kinetic theories that describe the simultaneous assembly of viral proteins and encapsidation of cargo.  When applied to the encapsidation of charge-functionalized nanoparticles, the theories predicts a transition from inefficient core encapsidation to nearly 100\% incorporation efficiency as the functionalized charge density is increased beyond a threshold value, and the estimated light scatter signal shows an initial rapid increase followed by slow rise to saturation, as seen in experiments. The predicted increase in incorporation efficiency with increasing charge density is sharper than seen in experiments, which could indicate the presence of kinetic traps that are not accounted for in the present theory; comparison of theory predictions with experimental data collected with a wide range of control parameters will be important to assess this possibility. The trends predicted here for varying surface charge, subunit concentration, and capsid-nanoparticle stoichiometric ratio could guide the design of experiments that identify fundamental principles and/or additional complexities for simultaneous assembly and cargo encapsulation.

{\bf Acknowledgments}  I am indebted to Bogdan Dragnea and the Dragnea group for insightful conversations and for providing me with experimental data, and to Oren Elrad for Fig~\ref{fig:adaptiveAssemblyLS}.  Funding was provided by Award Number R01AI080791 from the National Institute Of Allergy And Infectious Diseases and by the National Science Foundation through the Brandeis Materials Research Science and Engineering Center (MRSEC).

\appendix

\section{}
\label{sec:appB}

 \begin{figure} [bt]
\epsfig{file=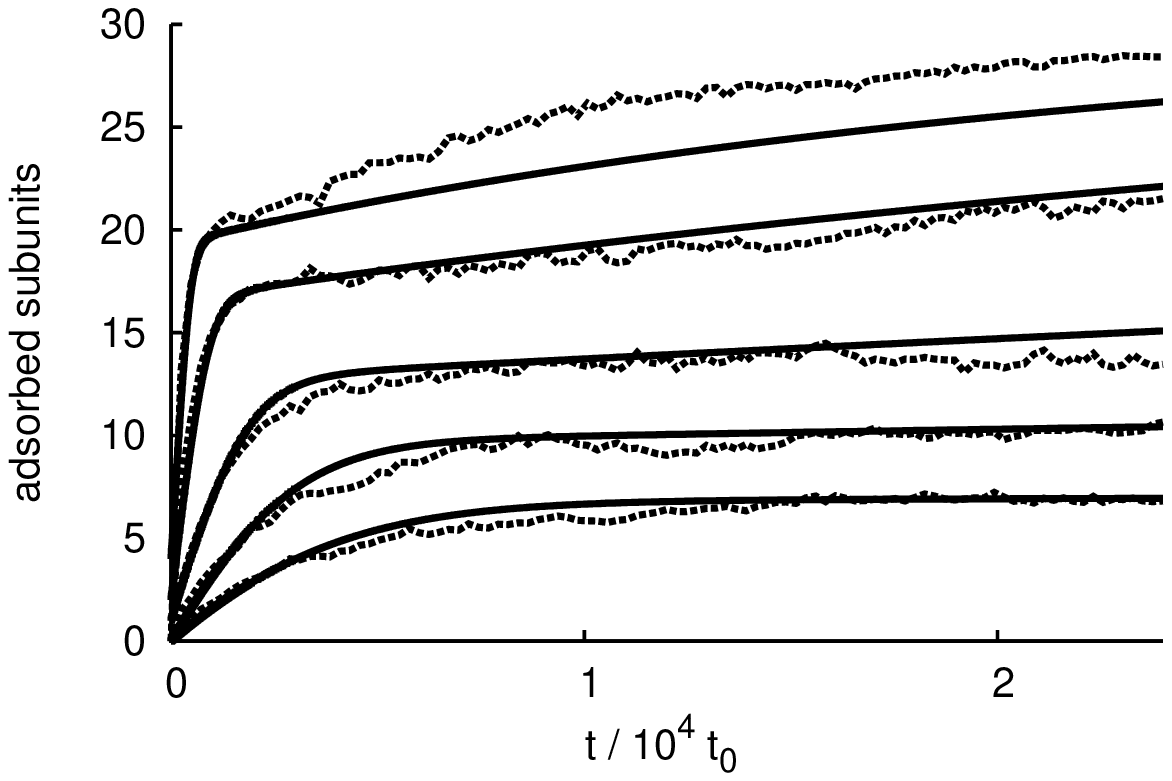,width = .5\textwidth}
\epsfig{file=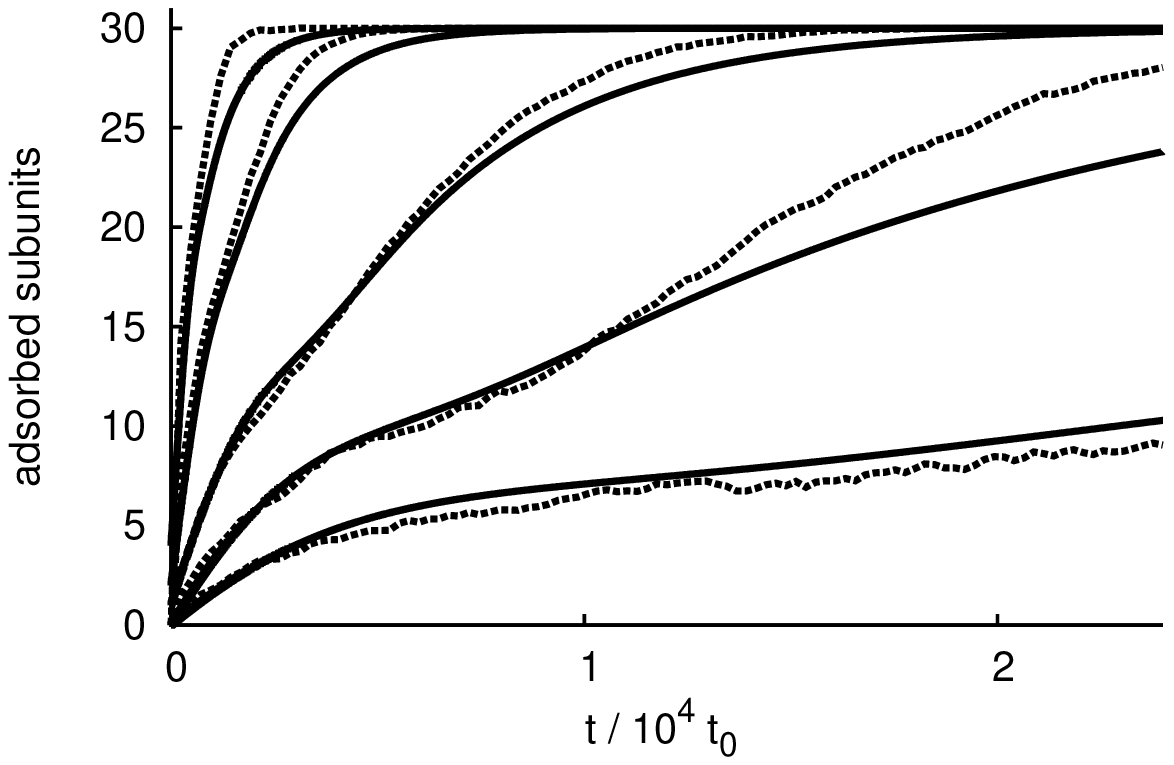,width = .5\textwidth}
\caption{\label{fig:ktCompareSims}
The kinetic theory predictions (smooth lines) and simulation results (noisy lines) for the time dependence of adsorbed and assembled subunits, $n+m$, are shown for the neutral subunit model (Eq.~\ref{eq:Gcs}). The simulations consider T=1 capsids and commensurate cores, so a complete capsid is comprised of $N=30$ dimer subunits. Curves at increasing height correspond to reduced subunit densities of $\CS a^3=$ 2.04, 4.07, 8.14, 20.4, 40.7, with a surface free energy of $\ec=7$ and subunit-subunit binding energies of (a) $\eb=9.0$  and (b) $\eb=11.0$
}
\end{figure}

In this appendix we compare predictions of the kinetic theory for the time dependence of the number of adsorbed particles, $n+m$,  to results presented in Ref.~\cite{Hagan2008}  from a computational model that represents the assembly of T=1 capsid-like objects around a commensurate nanoparticle.  This comparison elucidates the effect of approximations used in the kinetic theory. In particular, in rate equation approaches that assume a single reaction pathway as described above, or multiple reaction pathways \cite{Endres2005,Sweeney2008, Zhang2006}, and even binding between large intermediates \cite{Sweeney2008,Zhang2006}, only intermediates that are consistent with partially assembled capsids are considered.   The simulations, on the other hand, explicitly track the spatial coordinates of subunits and thereby make no a priori assumptions about assembly pathways.  Because we can either calculate parameters or estimate them from independent simulations,  this comparison is a stringent test of the validity of these approximations. It also tests the utility of using the kinetic theory to describe experimental adsorption results.  The procedure by which we determine parameters for the kinetic theory that correspond to particular sets of simulation parameters is described below, as well as modifications to the theory in order to represent neutral subunits (or high salt concentrations).  We do not consider here the simulations reported in  Fig.~\ref{fig:adaptiveAssemblyLS} and Ref.~\cite{Elrad2008}, for which the nanoparticle is not commensurate with the preferred capsid morphology.

Adsorption kinetics predicted by the kinetic theory and observed in simulations are shown in Fig.~\ref{fig:ktCompareSims} for several subunit concentrations and subunit-subunit binding energies for an adsorption energy of $\ec=7$.  The agreement between the kinetic theory and simulation results is surprisingly good, considering that no parameters were adjusted to fit this data and, as discussed below, relatively crude estimates are used for the subunit-subunit binding rate constants $f$ and binding entropy $\ssb$.

 The computational model considered in Ref.~\cite{Hagan2008} consists of rigid subunits,  for which excluded volume interactions are modeled by spherically symmetric repulsive forces, and complementary subunit-subunit interactions that drive assembly are modeled by directional attractions.  The lowest energy states in the model correspond to "capsids", which consist of multiples of 60 monomers in a shell with icosahedral symmetry.  The parameters of the model are the energy associated with the attractive potential, $\eb$ , which is measured in units of the thermal energy  $\kt$, and the specificity of the directional attractions, which is controlled by the angular parameters  $\tmax$ and $\pmax$.  Capsid subunits also experience short ranged isotropic interactions with a rigid sphere placed at the center of the simulation box; these interactions are minimized when capsid subunits are adjacent to the surface of the sphere.  Subunit positions and orientations are propagated according to overdamped Brownian dynamics, with the unit of time $t_0 = a^2/48D$, where D is the subunit diffusion coefficient.  Full details of the model are given in Ref.~\cite{Hagan2008}

\emph{Free energies for neutral subunits with no internal structure or high salt concentrations.}  The free energies for partial capsids on core surfaces are modified from the expressions given in Section~\ref{sec:coreFreeEnergy} for the case of neutral subunits or high salt concentrations, in which case interactions are limited to excluded volume, directional attractions that drive assembly, and favorable core-subunit interactions.    The free energy of a core with $n$ adsorbed but unassembled subunits and an adsorbed intermediate of size $m$ is written as
\begin{equation}
\Gcs_{n,m}=(n+m)\ec-T\Smix_{m,n}-T\Sad{n,m}
\label{eq:Gcs}
\end{equation}
where $\eb$ is the core-subunit surface free energy strength.

The second term, $\Smix$ accounts for the entropy for two-dimensional motions of adsorbed subunits on the core surface.  A simple approach would  assume Langmuir adsorption, but we find that much more accurate results are obtained by integrating an empirical formula for the chemical potential of a fluid of hard disks \cite{Siepmann1992} because the Langmuir model overestimates subunit mixing entropy at high surface coverages:
\begin{equation}
\Smix(\eta)/k_\text{B}=\int_0^{\eta}d \eta \left[-\frac{7}{8}\log(1-\eta)+\frac{2 \eta}{1-\eta}+\frac{9\eta}{8(1-\eta)^2}\right]
\label{eq:harddisk}
\end{equation}
with the packing fraction $\eta=(n+m)a^2/[16 \pi (\Rc+a/2)^2]$.

The last term in Eq.~\ref{eq:Gcs}, $\Sad$, is the entropy penalty for subunits to be localized a smaller distance from the core surface than the subunit diameter, which we obtained by numerically integrating the partition function for an adsorbed subunit when comparing the kinetic theory to simulations.

Calculating the time dependence of adsorbed and assembled subunits requires choosing values for several theoretical parameters; here we describe the parameter values and how they were chosen.
\begin{enumerate}

\item Subunit binding entropy penalty.  Eq.~\ref{eq:Gi} includes a term $\ssb$ for the entropy loss (in addition to the subunit mixing entropy included in $\gb$) for a subunit to bind to a capsid that accounts for translational restrictions on scales smaller than the subunit diameter $a$ and rotational restrictions \cite{Hagan2006}; the entropy penalty should increase only slightly when a subunit changes from one to more than one bond.  The subunit-subunit binding entropy penalty is calculated as described in Ref.~\cite{Hagan2006}.  There is a typo in Eq. (15) of that reference; the correct formula is
\begin{equation}
\ssb \approx-\frac{3}{2}\log \left.\frac{\beta \partial^2u_{\text{att}}(r)}{\partial r^2}\right|_{r=0} -\frac{1}{2}\log\frac{(\beta \varepsilon_{\text{b}})^3\pi^7}{\theta_{\text{m}}^4\phi_{\text{m}}^2}
\label{eq:sb}
\end{equation}

\item Subunit binding rate constant $f$.  The assembly rate constant was chosen as $f=0.03$ (in the dimensionless units of Ref.~\cite{Hagan2008}) by comparing kinetic theory predictions to simulation results for the assembly of empty capsids.  The resulting fit (not shown) shows less agreement between theory and simulation results than we find for core simulations, perhaps because approximations such as an average assembly pathway without multimer binding are less significant when subunits are confined to the surface of a core.

\item  The effective surface concentration of adsorbed subunits is increased if subunits are confined within less than a molecular diameter $a$ by the core-subunit interaction, so we take $\rho^\text{surf}_{n,m}=\frac{n}{a^3 (N-m)}\exp\left(S_{\text{ad}}/k_{\text{B}}\right)$.  This factor ensures that entropy losses due to subunit localization are not counted twice.

\item Adsorption rate, $\kad$.  The rate of subunit adsorption to core surfaces in our simulations can be calculated from the Smoluchowski equation, with forces due to the adsorption potential explicitly included.  Instead, we use the diffusion limited rate for the zero force case, but take $\kad=4 \pi D R_\text{cut}$, where $R_\text{cut}=2.5 a$ is the subunit-core center of mass distance at the point when the interaction force becomes nonzero.  This choice was validated by comparing theory and simulation results for the time dependence of the number of adsorbed subunits in the case of no assembly, i.e. $\eb = 0$. Since the excluded volume of adsorbed subunits is accounted for in Eq.~\ref{eq:harddisk} we set the adsorption blocking factor $\phi_{n,m}=1$ (see Eqs.~\ref{eq:ktCores}).  This approach slightly over estimates the adsorption rate at moderate coverage, but yields the proper equilibrium surface density if there is no assembly. Because of the definition of the unit time $t_0$, the subunit diffusion constant is $D=1/48$.
\end{enumerate}

 We find that the agreement in most cases is surprisingly good, considering the degree of error in estimating $\ssb$ and $f$, and the approximation that there is only one reaction pathway.

\section{}
\label{sec:appA}
In this appendix we summarize the application of the method of Scheutjens and Fleer \cite{Scheutjens1979} to polyelectrolytes.  Our presentation closely follows that given in Bohmer et al. \cite{Bohmer1990}, except there is a different geometry, the Flory-Huggins parameter for interactions between polymer segments is set to $\chi=0$, and there are additional terms in the free energy equations.

We consider a lattice around an impenetrable sphere with radius $\Rc$, contacting bulk solution in layer $M+1$, with $M$ large enough that the presence of the surface is negligible in layer $z=M$.  All quantities are assumed uniform within a layer $z$. Fixed numbers of polyelectrolytes and TEG molecules are grafted in layers $\zg=\Rcap/l=26$ and $\zc=\Rc/l+1=17$, respectively, with the lattice dimension $l=0.35$ nm.


In the calculation, a segment of species $i$ is subjected to a potential field $u_i(z)$, which is given by (relative to the bulk solution)
\begin{equation}
u_i(z)=\up(z)+v_i  e \Psi(z)
\label{eq:u}
\end{equation}
where the term $\up(z)$ represents the hard-core interaction, which is the same for every species and ensures that the total volume fraction for all species sums to unity in layer $z$.  The second term on the right-hand side accounts for electrostatic interactions, with the valency $v_i$ for a segment of species $i$, $e$ the charge on an electron, and  $\Psi(z)$ the electrostatic potential in layer $z$.
The statistical weight of a segment for species $i$ in layer $z$, relative to the bulk solution, is given by the segment weighting factor
\begin{equation}
G_i(z)=\exp(-u_i(z)/\kt)
\label{eq:GSF}
\end{equation}
so that the volume fraction of a free monomer of type $i$ is given by $\phi_i(z)=\phib_i G_i(z)$.

The volume fractions for segments in chain molecules are calculated with a segment distribution function $G_i(z,s|1)$, which gives the statistical weight of a chain conformation starting with a segment 1, located anywhere in the lattice, and ending in layer $z$ after $s-1$ steps.  The segment distribution function is calculated from a recurrence relation:
\begin{eqnarray}
\lefteqn{G_i(z,1|1)=G_i(z)} \nonumber \\
\lefteqn{G_i(z,s|1)=G_i(z)[\lambda_{-1} G_i(z-1,s-1|1)} \nonumber \\
              &  &+\lambda_0(G_i(z,s-1|1)+ \lambda_1 G_i(z+1,s-1|1)]
\label{eq:Greens}
\end{eqnarray}
with $\lambda_{-1}$, $\lambda_0$, and $\lambda_1$ the fraction of contacts for a lattice site in layer $z$ with respective layers $z-1$,$z$, and $z+1$ (see Ref.~\cite{Vanlent1989}).  The statistical weight $G_i(z,s|r)$ of a chain conformation that starts at segment $r$ and ends with segment $s$  in layer $z$  is calculated in the same way.   The segment distribution functions for grafted molecules are modified to constrain segment $1$ to begin in the grafted layer as described in Ref.~\cite{Cosgrove1987}.  To account for impenetrable surfaces, $u_i(z)=\infty$ for $z<\zc$ for all species, and $u_i(z)=\infty$ for $z>\zg$ for polyelectrolyte and TEG molecules.

The volume fraction of segment $s$ in a chain of $r$ segments is determined from the joint probability that a chain conformation starts at segment 1 and ends at segment $s$ in layer $z$, and a chain conformation starts at segment $r$ and ends at segment $s$ in layer $z$:
\begin{equation}
\phi_i(z,s)=C_i G_i(z,s|1)G_i(z,s|r)/G_i(z)
\label{eq:phi}
\end{equation}
where the probabilities are divided by $G_i(z)$ to avoid double counting of segment $s$ and $C_i$ is a normalization constant.  If molecules of species $i$ are free to exchange with the bulk solution, $C_i$ is determined by the fact that all segment weighting factors $G_i=1$ in bulk so that
\begin{equation}
C_i=\phib_i/r_i
\label{eq:c1}.
\end{equation}
If the total amount of molecules of species $i$ is fixed (i.e. for a grafted brush) $C_i$ is determined from
\begin{equation}
C_i=\theta_i/[r_i G_i(r|1)]
\label{eq:c2}
\end{equation}
with the total amount of molecules of species $i$ calculated from
\begin{equation}
\theta_i=\sum_{z=1}^M L(z)\phi_i(z)
\label{eq:theta}
\end{equation}
and the chain weighting factor $G_i(r|1)=\sum_{z=1}^M L(z) G_i(z,r|1)$, which gives the statistical weight for a chain of type $i$ to be found anywhere in the lattice. The number of lattice sites $L(z)$ depends on the layer $z$ because of curvature and is calculated in Ref.~\cite{Vanlent1989}.

\emph{The electrostatic potential.}  Electrostatics are accounted for with a multi-Stern-layer model in which all charges within a lattice layer are located on the plane at the center of the layer, with no charge outside of that plane.  The plane charge density in layer $z$, $\sigma(z)$ is calculated from a sum over all species
\begin{equation}
\sigma(z)=\sum_i \alpha_i(z) v_i e \phi_i(z)  /\as
\label{eq:sigma}
\end{equation}
where we set the cross-sectional area per lattice site $\as=l^2$ with $l$ the distance between layers. For weak acid groups the degree of dissociation $\alpha_i(z)$ is given by\cite{Bohmer1991}
\begin{equation}
\alpha_i(z)=\frac{K_i \langle \phiWater(z)\rangle}{\langle \phiHplus(z)\rangle + K_i \langle \phiWater(z) \rangle}
\label{eq:alpha}
\end{equation}
with the dimensionless dissociation parameter $K_i$ related to the dissociation constant $K_\text{D}$ by $K_i=K_\text{D}/c_i$ with $c_i$ the molarity of pure segments for species $i$.

The electrostatic potential with respect to bulk solution at layer $M$ is calculated through electroneutrality \cite{Bohmer1990} and the potential in each layer is calculated from Gauss' law, for spherical layers it is  \cite{Bohmer1991}
\begin{eqnarray}
\lefteqn{\Psi(z+1)= \Psi(z)- \frac{l}{8 \pi l} \left[\frac{1}{\epsilon(z)(z-1/2)z}\:+ \right.} \nonumber \\
          &   &\left. \frac{1}{\epsilon(z+1)z (z+1/2)} \right]
          \sum_{\zp=1}^{z}L(\zp) \sigma(\zp)
\label{eq:Psi}
\end{eqnarray}
with $\epsilon(z)$ the volume fraction averaged dielectric permittivity for layer $z$.

\emph{Free energies.}  The free energy is obtained from the canonical partition function for the lattice system \cite{Scheutjens1979, Evers1990, Bohmer1990}
\begin{eqnarray}
\lefteqn{A =  \kt \sum_i \frac{\theta_i}{r_i} \ln(r_i C_i)} \nonumber \\
          &     &  -\sum_i \sum_z L(z) \phi_i(z) u_i(z) + U_\text{EL}  + F_\text{diss}
\label{eq:freeEnergy}
\end{eqnarray}
with the electrostatic energy for the case of a fixed surface charge given by
\begin{equation}
\frac{U_\text{EL}}{\kt}=\frac{1}{2}\sum_{z=1}^{M+1} L(z) \sigma(z)\Psi(z)
\label{eq:Uel}.
\end{equation}
The final term in  Eq.~\ref{eq:freeEnergy} was omitted in Ref.~\cite{Bohmer1990}, but is necessary to describe the free energy due to dissociation of acid groups
\begin{eqnarray}
\lefteqn{\frac{F_\text{diss}}{\kt} = \sum_{z=1}^{M} L(z) \sum_{a} \phi_{a} |v_a| \bigg[ \alpha_a(z) \log \frac{\phib_{\text{H}_3\text{O}^{+}}}{K_a}  + }  \\
& & \alpha_a(z) \log \alpha_a(z) +  (1-\alpha_a(z)) \log (1-\alpha_a(z)) \bigg] \nonumber
\label{eq:Fdiss}
\end{eqnarray}
where the sum extends over the weak acid species $a$.

The Gibbs excess free energy $A_\text{EX}$ is obtained from
\begin{equation}
 A_\text{EX} = A - \sum_{\ip}\frac{\theta_{\ip}}{r_{\ip}} \mu_\ip
\label{eq:Aexcess}
\end{equation}
where the sum extends over the species $\ip$ that are in equilibrium with the bulk solution.  The Flory-Huggins equation for the chemical potential $\mu_i$ of a mixture of chain molecules is \cite{Bohmer1990} (with $\chi=0$)
\begin{equation}
\frac{\mu_i}{\kt}=\ln \phib_i + 1-r_i\sum_j\frac{\phib_j}{r_j}
\label{eq:mu}.
\end{equation}
For calculations with a fixed density of grafted polymers, the free energy includes additional terms, and the free energy $\Fsf$ (per area at the interior surface of the capsid) is
\begin{eqnarray}
\lefteqn{\frac{\Fsf(\{\sg\}) L(\zg)a_\text{s}}{\kt} = \frac{A}{\kt}  + \sum_g L(z_g) \left[ \sg \log \sg  \right. } \nonumber \\
& &  \left. + (1-\sg)\log(1-\sg) + \sg(r_g-1) \right]
\label{eq:Fsf}
\end{eqnarray}
where the sum is over the $g\in \{i\} \setminus \{\ip\}$ grafted species and $\sg=\theta_g/(L(z_g) r_g)$ is the grafting density (per lattice site) of species $g$. The first two terms in the sum account for fixed locations of grafted (or assembled in a capsid) segments \cite{Lent1990} and  the third adjusts the polymer reference state from the pure liquid to solvated polymer in infinite dilution (the last two terms of  Eq.~\ref{eq:mu} with $\phib=0$).  With this choice of reference state, the excess free energy $\Fex$ as a function of the amount of adsorbed polyelectrolyte is
\begin{equation}
\Fex(\sigma_\text{p})=\Fsf(\sigma_\text{p})+\kt \sigma_\text{p} \log(a^3 \CS)
\label{eq:fexc}
\end{equation}
where $a^3$ is the standard state volume and we have suppressed the dependence on $\sigma_\text{TEG}$, which is constant for our calculations.

We consider 9 molecular species, $\text{H}_2\text{O}$, $\text{OH}^{-}$, $\text{H}_3\text{O}^{+}$, monovalent positive salt ions, monovalent negative salt ions, divalent positive salt ions, polyelectrolyte, neutral TEG, and acidic TEG end groups.  Following Bohmer et al.\cite{Bohmer1990} the number of species is reduced to 8 by treating water as a weak electrolyte with valence $-1$ and dissociation constant $pK_\text{w}=14-2 \log(10^3 N_\text{A} l^3)$ ( Eq.~\ref{eq:alpha} for $\phi_\text{w}=1$, $N_\text{A}$ Avogadro's number, and $l^3$ the volume of a lattice site).


\begin{thebibliography}{92}
\expandafter\ifx\csname natexlab\endcsname\relax\def\natexlab#1{#1}\fi
\expandafter\ifx\csname bibnamefont\endcsname\relax
  \def\bibnamefont#1{#1}\fi
\expandafter\ifx\csname bibfnamefont\endcsname\relax
  \def\bibfnamefont#1{#1}\fi
\expandafter\ifx\csname citenamefont\endcsname\relax
  \def\citenamefont#1{#1}\fi
\expandafter\ifx\csname url\endcsname\relax
  \def\url#1{\texttt{#1}}\fi
\expandafter\ifx\csname urlprefix\endcsname\relax\def\urlprefix{URL }\fi
\providecommand{\bibinfo}[2]{#2}
\providecommand{\eprint}[2][]{\url{#2}}

\bibitem[{\citenamefont{Gupta et~al.}(2005)\citenamefont{Gupta, Levchenko, and
  Torchilin}}]{Gupta2005}
\bibinfo{author}{\bibfnamefont{B.}~\bibnamefont{Gupta}},
  \bibinfo{author}{\bibfnamefont{T.~S.} \bibnamefont{Levchenko}},
  \bibnamefont{and} \bibinfo{author}{\bibfnamefont{V.~P.}
  \bibnamefont{Torchilin}}, \bibinfo{journal}{Adv. Drug Delivery Rev.}
  \textbf{\bibinfo{volume}{57}}, \bibinfo{pages}{637} (\bibinfo{year}{2005}).

\bibitem[{\citenamefont{Garcea and Gissmann}(2004)}]{Garcea2004}
\bibinfo{author}{\bibfnamefont{R.~L.} \bibnamefont{Garcea}} \bibnamefont{and}
  \bibinfo{author}{\bibfnamefont{L.}~\bibnamefont{Gissmann}},
  \bibinfo{journal}{Curr. Opin. Biotechnol.} \textbf{\bibinfo{volume}{15}},
  \bibinfo{pages}{513} (\bibinfo{year}{2004}).

\bibitem[{\citenamefont{Dietz and Bahr}(2004)}]{Dietz2004}
\bibinfo{author}{\bibfnamefont{G.~P.~H.} \bibnamefont{Dietz}} \bibnamefont{and}
  \bibinfo{author}{\bibfnamefont{M.}~\bibnamefont{Bahr}},
  \bibinfo{journal}{Mol. Cell. Neurosci.} \textbf{\bibinfo{volume}{27}},
  \bibinfo{pages}{85} (\bibinfo{year}{2004}).

\bibitem[{\citenamefont{Soto et~al.}(2006)\citenamefont{Soto, Blum, Vora,
  Lebedev, Meador, Won, Chatterji, Johnson, and Ratna}}]{Soto2006}
\bibinfo{author}{\bibfnamefont{C.~M.} \bibnamefont{Soto}},
  \bibinfo{author}{\bibfnamefont{A.~S.} \bibnamefont{Blum}},
  \bibinfo{author}{\bibfnamefont{G.~J.} \bibnamefont{Vora}},
  \bibinfo{author}{\bibfnamefont{N.}~\bibnamefont{Lebedev}},
  \bibinfo{author}{\bibfnamefont{C.~E.} \bibnamefont{Meador}},
  \bibinfo{author}{\bibfnamefont{A.~P.} \bibnamefont{Won}},
  \bibinfo{author}{\bibfnamefont{A.}~\bibnamefont{Chatterji}},
  \bibinfo{author}{\bibfnamefont{J.~E.} \bibnamefont{Johnson}},
  \bibnamefont{and} \bibinfo{author}{\bibfnamefont{B.~R.} \bibnamefont{Ratna}},
  \bibinfo{journal}{J. Am. Chem. Soc.} \textbf{\bibinfo{volume}{128}},
  \bibinfo{pages}{5184} (\bibinfo{year}{2006}).

\bibitem[{\citenamefont{Sapsford et~al.}(2006)\citenamefont{Sapsford, Soto,
  Blum, Chatterji, Lin, Johnson, Ligler, and Ratna}}]{Sapsford2006}
\bibinfo{author}{\bibfnamefont{K.~E.} \bibnamefont{Sapsford}},
  \bibinfo{author}{\bibfnamefont{C.~M.} \bibnamefont{Soto}},
  \bibinfo{author}{\bibfnamefont{A.~S.} \bibnamefont{Blum}},
  \bibinfo{author}{\bibfnamefont{A.}~\bibnamefont{Chatterji}},
  \bibinfo{author}{\bibfnamefont{T.~W.} \bibnamefont{Lin}},
  \bibinfo{author}{\bibfnamefont{J.~E.} \bibnamefont{Johnson}},
  \bibinfo{author}{\bibfnamefont{F.~S.} \bibnamefont{Ligler}},
  \bibnamefont{and} \bibinfo{author}{\bibfnamefont{B.~R.} \bibnamefont{Ratna}},
  \bibinfo{journal}{Biosens. Bioelectron.} \textbf{\bibinfo{volume}{21}},
  \bibinfo{pages}{1668} (\bibinfo{year}{2006}).

\bibitem[{\citenamefont{Boldogkoi et~al.}(2004)\citenamefont{Boldogkoi, Sik,
  Denes, Reichart, Toldi, Gerendai, Kovacs, and Palkovits}}]{Boldogkoi2004}
\bibinfo{author}{\bibfnamefont{Z.}~\bibnamefont{Boldogkoi}},
  \bibinfo{author}{\bibfnamefont{A.}~\bibnamefont{Sik}},
  \bibinfo{author}{\bibfnamefont{A.}~\bibnamefont{Denes}},
  \bibinfo{author}{\bibfnamefont{A.}~\bibnamefont{Reichart}},
  \bibinfo{author}{\bibfnamefont{J.}~\bibnamefont{Toldi}},
  \bibinfo{author}{\bibfnamefont{I.}~\bibnamefont{Gerendai}},
  \bibinfo{author}{\bibfnamefont{K.~J.} \bibnamefont{Kovacs}},
  \bibnamefont{and}
  \bibinfo{author}{\bibfnamefont{M.}~\bibnamefont{Palkovits}},
  \bibinfo{journal}{Prog. Neurobiol.} \textbf{\bibinfo{volume}{72}},
  \bibinfo{pages}{417} (\bibinfo{year}{2004}).

\bibitem[{\citenamefont{Dragnea et~al.}(2003)\citenamefont{Dragnea, Chen, Kwak,
  Stein, and Kao}}]{Dragnea2003}
\bibinfo{author}{\bibfnamefont{B.}~\bibnamefont{Dragnea}},
  \bibinfo{author}{\bibfnamefont{C.}~\bibnamefont{Chen}},
  \bibinfo{author}{\bibfnamefont{E.~S.} \bibnamefont{Kwak}},
  \bibinfo{author}{\bibfnamefont{B.}~\bibnamefont{Stein}}, \bibnamefont{and}
  \bibinfo{author}{\bibfnamefont{C.~C.} \bibnamefont{Kao}},
  \bibinfo{journal}{J. Am. Chem. Soc.} \textbf{\bibinfo{volume}{125}},
  \bibinfo{pages}{6374} (\bibinfo{year}{2003}).

\bibitem[{\citenamefont{Chatterji et~al.}(2005)\citenamefont{Chatterji, Ochoa,
  Ueno, Lin, and Johnson}}]{Chatterji2005}
\bibinfo{author}{\bibfnamefont{A.}~\bibnamefont{Chatterji}},
  \bibinfo{author}{\bibfnamefont{W.~F.} \bibnamefont{Ochoa}},
  \bibinfo{author}{\bibfnamefont{T.}~\bibnamefont{Ueno}},
  \bibinfo{author}{\bibfnamefont{T.~W.} \bibnamefont{Lin}}, \bibnamefont{and}
  \bibinfo{author}{\bibfnamefont{J.~E.} \bibnamefont{Johnson}},
  \bibinfo{journal}{Nano Lett.} \textbf{\bibinfo{volume}{5}},
  \bibinfo{pages}{597} (\bibinfo{year}{2005}).

\bibitem[{\citenamefont{Falkner et~al.}(2005)\citenamefont{Falkner, Turner,
  Bosworth, Trentler, Johnson, Lin, and Colvin}}]{Falkner2005}
\bibinfo{author}{\bibfnamefont{J.~C.} \bibnamefont{Falkner}},
  \bibinfo{author}{\bibfnamefont{M.~E.} \bibnamefont{Turner}},
  \bibinfo{author}{\bibfnamefont{J.~K.} \bibnamefont{Bosworth}},
  \bibinfo{author}{\bibfnamefont{T.~J.} \bibnamefont{Trentler}},
  \bibinfo{author}{\bibfnamefont{J.~E.} \bibnamefont{Johnson}},
  \bibinfo{author}{\bibfnamefont{T.~W.} \bibnamefont{Lin}}, \bibnamefont{and}
  \bibinfo{author}{\bibfnamefont{V.~L.} \bibnamefont{Colvin}},
  \bibinfo{journal}{J. Am. Chem. Soc.} \textbf{\bibinfo{volume}{127}},
  \bibinfo{pages}{5274} (\bibinfo{year}{2005}).

\bibitem[{\citenamefont{Flynn et~al.}(2003)\citenamefont{Flynn, Lee, Peelle,
  and Belcher}}]{Flynn2003}
\bibinfo{author}{\bibfnamefont{C.~E.} \bibnamefont{Flynn}},
  \bibinfo{author}{\bibfnamefont{S.~W.} \bibnamefont{Lee}},
  \bibinfo{author}{\bibfnamefont{B.~R.} \bibnamefont{Peelle}},
  \bibnamefont{and} \bibinfo{author}{\bibfnamefont{A.~M.}
  \bibnamefont{Belcher}}, \bibinfo{journal}{Acta Mater.}
  \textbf{\bibinfo{volume}{51}}, \bibinfo{pages}{5867} (\bibinfo{year}{2003}).

\bibitem[{\citenamefont{Douglas and Young}(1998)}]{Douglas1998}
\bibinfo{author}{\bibfnamefont{T.}~\bibnamefont{Douglas}} \bibnamefont{and}
  \bibinfo{author}{\bibfnamefont{M.}~\bibnamefont{Young}},
  \bibinfo{journal}{Nature} \textbf{\bibinfo{volume}{393}},
  \bibinfo{pages}{152} (\bibinfo{year}{1998}).

\bibitem[{\citenamefont{Douglas and Young}(2006)}]{Douglas2006}
\bibinfo{author}{\bibfnamefont{T.}~\bibnamefont{Douglas}} \bibnamefont{and}
  \bibinfo{author}{\bibfnamefont{M.}~\bibnamefont{Young}},
  \bibinfo{journal}{Science} \textbf{\bibinfo{volume}{312}},
  \bibinfo{pages}{873(3)} (\bibinfo{year}{2006}).

\bibitem[{\citenamefont{Johnson
  et~al.}(2004{\natexlab{a}})\citenamefont{Johnson, Tang, Johnson, and
  Ball}}]{Johnson2004}
\bibinfo{author}{\bibfnamefont{K.~N.} \bibnamefont{Johnson}},
  \bibinfo{author}{\bibfnamefont{L.}~\bibnamefont{Tang}},
  \bibinfo{author}{\bibfnamefont{J.~E.} \bibnamefont{Johnson}},
  \bibnamefont{and} \bibinfo{author}{\bibfnamefont{L.~A.} \bibnamefont{Ball}},
  \bibinfo{journal}{J. Virol.} \textbf{\bibinfo{volume}{78}},
  \bibinfo{pages}{11371} (\bibinfo{year}{2004}{\natexlab{a}}).

\bibitem[{\citenamefont{Fox et~al.}(1994)\citenamefont{Fox, Johnson, and
  Young}}]{Fox1994}
\bibinfo{author}{\bibfnamefont{J.~M.} \bibnamefont{Fox}},
  \bibinfo{author}{\bibfnamefont{J.~E.} \bibnamefont{Johnson}},
  \bibnamefont{and} \bibinfo{author}{\bibfnamefont{M.~J.} \bibnamefont{Young}},
  \bibinfo{journal}{Seminars in Virology} \textbf{\bibinfo{volume}{5}},
  \bibinfo{pages}{51} (\bibinfo{year}{1994}).

\bibitem[{\citenamefont{Valegard et~al.}(1997)\citenamefont{Valegard, Murray,
  Stonehouse, vandenWorm, Stockley, and Liljas}}]{Valegard1997}
\bibinfo{author}{\bibfnamefont{K.}~\bibnamefont{Valegard}},
  \bibinfo{author}{\bibfnamefont{J.~B.} \bibnamefont{Murray}},
  \bibinfo{author}{\bibfnamefont{N.~J.} \bibnamefont{Stonehouse}},
  \bibinfo{author}{\bibfnamefont{S.}~\bibnamefont{vandenWorm}},
  \bibinfo{author}{\bibfnamefont{P.~G.} \bibnamefont{Stockley}},
  \bibnamefont{and} \bibinfo{author}{\bibfnamefont{L.}~\bibnamefont{Liljas}},
  \bibinfo{journal}{J. Mol. Biol.} \textbf{\bibinfo{volume}{270}},
  \bibinfo{pages}{724} (\bibinfo{year}{1997}).

\bibitem[{\citenamefont{Johnson
  et~al.}(2004{\natexlab{b}})\citenamefont{Johnson, Tang, Johnson, and
  Ball}}]{Johnson2004b}
\bibinfo{author}{\bibfnamefont{K.~N.} \bibnamefont{Johnson}},
  \bibinfo{author}{\bibfnamefont{L.}~\bibnamefont{Tang}},
  \bibinfo{author}{\bibfnamefont{J.~E.} \bibnamefont{Johnson}},
  \bibnamefont{and} \bibinfo{author}{\bibfnamefont{L.~A.} \bibnamefont{Ball}},
  \bibinfo{journal}{J. Virol.} \textbf{\bibinfo{volume}{78}},
  \bibinfo{pages}{11371} (\bibinfo{year}{2004}{\natexlab{b}}).

\bibitem[{\citenamefont{Tihova et~al.}(2004)\citenamefont{Tihova, Dryden, Le,
  Harvey, Johnson, Yeager, and Schneemann}}]{Tihova2004}
\bibinfo{author}{\bibfnamefont{M.}~\bibnamefont{Tihova}},
  \bibinfo{author}{\bibfnamefont{K.~A.} \bibnamefont{Dryden}},
  \bibinfo{author}{\bibfnamefont{T.~V.~L.} \bibnamefont{Le}},
  \bibinfo{author}{\bibfnamefont{S.~C.} \bibnamefont{Harvey}},
  \bibinfo{author}{\bibfnamefont{J.~E.} \bibnamefont{Johnson}},
  \bibinfo{author}{\bibfnamefont{M.}~\bibnamefont{Yeager}}, \bibnamefont{and}
  \bibinfo{author}{\bibfnamefont{A.}~\bibnamefont{Schneemann}},
  \bibinfo{journal}{J. Virol.} \textbf{\bibinfo{volume}{78}},
  \bibinfo{pages}{2897} (\bibinfo{year}{2004}).

\bibitem[{\citenamefont{Krol et~al.}(1999)\citenamefont{Krol, Olson, Tate,
  Johnson, Baker, and Ahlquist}}]{Krol1999}
\bibinfo{author}{\bibfnamefont{M.~A.} \bibnamefont{Krol}},
  \bibinfo{author}{\bibfnamefont{N.~H.} \bibnamefont{Olson}},
  \bibinfo{author}{\bibfnamefont{J.}~\bibnamefont{Tate}},
  \bibinfo{author}{\bibfnamefont{J.~E.} \bibnamefont{Johnson}},
  \bibinfo{author}{\bibfnamefont{T.~S.} \bibnamefont{Baker}}, \bibnamefont{and}
  \bibinfo{author}{\bibfnamefont{P.}~\bibnamefont{Ahlquist}},
  \bibinfo{journal}{Proc. Natl. Acad. Sci. U. S. A.}
  \textbf{\bibinfo{volume}{96}}, \bibinfo{pages}{13650} (\bibinfo{year}{1999}).

\bibitem[{\citenamefont{Stockley et~al.}(2007)\citenamefont{Stockley, Rolfsson,
  Thompson, Basnak, Francese, Stonehouse, Homans, and Ashcroft}}]{Stockley2007}
\bibinfo{author}{\bibfnamefont{P.~G.} \bibnamefont{Stockley}},
  \bibinfo{author}{\bibfnamefont{O.}~\bibnamefont{Rolfsson}},
  \bibinfo{author}{\bibfnamefont{G.~S.} \bibnamefont{Thompson}},
  \bibinfo{author}{\bibfnamefont{G.}~\bibnamefont{Basnak}},
  \bibinfo{author}{\bibfnamefont{S.}~\bibnamefont{Francese}},
  \bibinfo{author}{\bibfnamefont{N.~J.} \bibnamefont{Stonehouse}},
  \bibinfo{author}{\bibfnamefont{S.~W.} \bibnamefont{Homans}},
  \bibnamefont{and} \bibinfo{author}{\bibfnamefont{A.~E.}
  \bibnamefont{Ashcroft}}, \bibinfo{journal}{J. Mol. BIo.}
  \textbf{\bibinfo{volume}{369}}, \bibinfo{pages}{541} (\bibinfo{year}{2007}).

\bibitem[{\citenamefont{Toropova et~al.}(2008)\citenamefont{Toropova, Basnak,
  Twarock, Stockley, and Ranson}}]{Toropova2008}
\bibinfo{author}{\bibfnamefont{K.}~\bibnamefont{Toropova}},
  \bibinfo{author}{\bibfnamefont{G.}~\bibnamefont{Basnak}},
  \bibinfo{author}{\bibfnamefont{R.}~\bibnamefont{Twarock}},
  \bibinfo{author}{\bibfnamefont{P.~G.} \bibnamefont{Stockley}},
  \bibnamefont{and} \bibinfo{author}{\bibfnamefont{N.~A.}
  \bibnamefont{Ranson}}, \bibinfo{journal}{J. Mol. Biol.}
  \textbf{\bibinfo{volume}{375}}, \bibinfo{pages}{824} (\bibinfo{year}{2008}).

\bibitem[{\citenamefont{Bancroft et~al.}(1969)\citenamefont{Bancroft, Hiebert,
  and Bracker}}]{Bancroft1969}
\bibinfo{author}{\bibfnamefont{J.~B.} \bibnamefont{Bancroft}},
  \bibinfo{author}{\bibfnamefont{E.}~\bibnamefont{Hiebert}}, \bibnamefont{and}
  \bibinfo{author}{\bibfnamefont{C.~E.} \bibnamefont{Bracker}},
  \bibinfo{journal}{Virology} \textbf{\bibinfo{volume}{39}},
  \bibinfo{pages}{924} (\bibinfo{year}{1969}).

\bibitem[{\citenamefont{Hu et~al.}(2008)\citenamefont{Hu, Zandi, Anavitarte,
  Knobler, and Gelbart}}]{Hu2008}
\bibinfo{author}{\bibfnamefont{Y.}~\bibnamefont{Hu}},
  \bibinfo{author}{\bibfnamefont{R.}~\bibnamefont{Zandi}},
  \bibinfo{author}{\bibfnamefont{A.}~\bibnamefont{Anavitarte}},
  \bibinfo{author}{\bibfnamefont{C.~M.} \bibnamefont{Knobler}},
  \bibnamefont{and} \bibinfo{author}{\bibfnamefont{W.~M.}
  \bibnamefont{Gelbart}}, \bibinfo{journal}{Biophys. J.}
  \textbf{\bibinfo{volume}{94}}, \bibinfo{pages}{1428} (\bibinfo{year}{2008}).

\bibitem[{\citenamefont{Sikkema et~al.}(2007)\citenamefont{Sikkema,
  Comellas-Aragones, Fokkink, Verduin, Cornelissen, and Nolte}}]{Sikkema2007}
\bibinfo{author}{\bibfnamefont{F.~D.} \bibnamefont{Sikkema}},
  \bibinfo{author}{\bibfnamefont{M.}~\bibnamefont{Comellas-Aragones}},
  \bibinfo{author}{\bibfnamefont{R.~G.} \bibnamefont{Fokkink}},
  \bibinfo{author}{\bibfnamefont{B.~J.~M.} \bibnamefont{Verduin}},
  \bibinfo{author}{\bibfnamefont{J.}~\bibnamefont{Cornelissen}},
  \bibnamefont{and} \bibinfo{author}{\bibfnamefont{R.~J.~M.}
  \bibnamefont{Nolte}}, \bibinfo{journal}{Org. Biomol. Chem.}
  \textbf{\bibinfo{volume}{5}}, \bibinfo{pages}{54} (\bibinfo{year}{2007}).

\bibitem[{\citenamefont{Sun et~al.}(2007)\citenamefont{Sun, DuFort, Daniel,
  Murali, Chen, Gopinath, Stein, De, Rotello, Holzenburg et~al.}}]{Sun2007}
\bibinfo{author}{\bibfnamefont{J.}~\bibnamefont{Sun}},
  \bibinfo{author}{\bibfnamefont{C.}~\bibnamefont{DuFort}},
  \bibinfo{author}{\bibfnamefont{M.~C.} \bibnamefont{Daniel}},
  \bibinfo{author}{\bibfnamefont{A.}~\bibnamefont{Murali}},
  \bibinfo{author}{\bibfnamefont{C.}~\bibnamefont{Chen}},
  \bibinfo{author}{\bibfnamefont{K.}~\bibnamefont{Gopinath}},
  \bibinfo{author}{\bibfnamefont{B.}~\bibnamefont{Stein}},
  \bibinfo{author}{\bibfnamefont{M.}~\bibnamefont{De}},
  \bibinfo{author}{\bibfnamefont{V.~M.} \bibnamefont{Rotello}},
  \bibinfo{author}{\bibfnamefont{A.}~\bibnamefont{Holzenburg}},
  \bibnamefont{et~al.}, \bibinfo{journal}{Proc. Natl. Acad. Sci. U. S. A.}
  \textbf{\bibinfo{volume}{104}}, \bibinfo{pages}{1354} (\bibinfo{year}{2007}).

\bibitem[{\citenamefont{Dixit et~al.}(2006)\citenamefont{Dixit, Goicochea,
  Daniel, Murali, Bronstein, De, Stein, Rotello, Kao, and Dragnea}}]{Dixit2006}
\bibinfo{author}{\bibfnamefont{S.~K.} \bibnamefont{Dixit}},
  \bibinfo{author}{\bibfnamefont{N.~L.} \bibnamefont{Goicochea}},
  \bibinfo{author}{\bibfnamefont{M.~C.} \bibnamefont{Daniel}},
  \bibinfo{author}{\bibfnamefont{A.}~\bibnamefont{Murali}},
  \bibinfo{author}{\bibfnamefont{L.}~\bibnamefont{Bronstein}},
  \bibinfo{author}{\bibfnamefont{M.}~\bibnamefont{De}},
  \bibinfo{author}{\bibfnamefont{B.}~\bibnamefont{Stein}},
  \bibinfo{author}{\bibfnamefont{V.~M.} \bibnamefont{Rotello}},
  \bibinfo{author}{\bibfnamefont{C.~C.} \bibnamefont{Kao}}, \bibnamefont{and}
  \bibinfo{author}{\bibfnamefont{B.}~\bibnamefont{Dragnea}},
  \bibinfo{journal}{Nano Lett.} \textbf{\bibinfo{volume}{6}},
  \bibinfo{pages}{1993} (\bibinfo{year}{2006}).

\bibitem[{\citenamefont{Chen et~al.}(2005)\citenamefont{Chen, Kwak, Stein, Kao,
  and Dragnea}}]{Chen2005}
\bibinfo{author}{\bibfnamefont{C.}~\bibnamefont{Chen}},
  \bibinfo{author}{\bibfnamefont{E.~S.} \bibnamefont{Kwak}},
  \bibinfo{author}{\bibfnamefont{B.}~\bibnamefont{Stein}},
  \bibinfo{author}{\bibfnamefont{C.~C.} \bibnamefont{Kao}}, \bibnamefont{and}
  \bibinfo{author}{\bibfnamefont{B.}~\bibnamefont{Dragnea}},
  \bibinfo{journal}{J. Nanosci. and Nanotech.} \textbf{\bibinfo{volume}{5}},
  \bibinfo{pages}{2029} (\bibinfo{year}{2005}).

\bibitem[{\citenamefont{Chen et~al.}(2006)\citenamefont{Chen, Daniel, Quinkert,
  De, Stein, Bowman, Chipman, Rotello, Kao, and Dragnea}}]{Chen2006}
\bibinfo{author}{\bibfnamefont{C.}~\bibnamefont{Chen}},
  \bibinfo{author}{\bibfnamefont{M.~C.} \bibnamefont{Daniel}},
  \bibinfo{author}{\bibfnamefont{Z.~T.} \bibnamefont{Quinkert}},
  \bibinfo{author}{\bibfnamefont{M.}~\bibnamefont{De}},
  \bibinfo{author}{\bibfnamefont{B.}~\bibnamefont{Stein}},
  \bibinfo{author}{\bibfnamefont{V.~D.} \bibnamefont{Bowman}},
  \bibinfo{author}{\bibfnamefont{P.~R.} \bibnamefont{Chipman}},
  \bibinfo{author}{\bibfnamefont{V.~M.} \bibnamefont{Rotello}},
  \bibinfo{author}{\bibfnamefont{C.~C.} \bibnamefont{Kao}}, \bibnamefont{and}
  \bibinfo{author}{\bibfnamefont{B.}~\bibnamefont{Dragnea}},
  \bibinfo{journal}{Nano Lett.} \textbf{\bibinfo{volume}{6}},
  \bibinfo{pages}{611} (\bibinfo{year}{2006}).

\bibitem[{\citenamefont{Chang et~al.}(2008)\citenamefont{Chang, Knobler,
  Gelbart, and Mason}}]{Chang2008}
\bibinfo{author}{\bibfnamefont{C.~B.} \bibnamefont{Chang}},
  \bibinfo{author}{\bibfnamefont{C.~M.} \bibnamefont{Knobler}},
  \bibinfo{author}{\bibfnamefont{W.~M.} \bibnamefont{Gelbart}},
  \bibnamefont{and} \bibinfo{author}{\bibfnamefont{T.~G.} \bibnamefont{Mason}},
  \bibinfo{journal}{Acs Nano} \textbf{\bibinfo{volume}{2}},
  \bibinfo{pages}{281} (\bibinfo{year}{2008}), ISSN \bibinfo{issn}{1936-0851}.

\bibitem[{\citenamefont{Zlotnick}(2007)}]{Zlotnick2007}
\bibinfo{author}{\bibfnamefont{A.}~\bibnamefont{Zlotnick}},
  \bibinfo{journal}{J. Mol. Biol.} \textbf{\bibinfo{volume}{366}},
  \bibinfo{pages}{14} (\bibinfo{year}{2007}).

\bibitem[{\citenamefont{Zlotnick}(1994)}]{Zlotnick1994}
\bibinfo{author}{\bibfnamefont{A.}~\bibnamefont{Zlotnick}},
  \bibinfo{journal}{J. Mol. Biol.} \textbf{\bibinfo{volume}{241}},
  \bibinfo{pages}{59} (\bibinfo{year}{1994}).

\bibitem[{\citenamefont{Endres and Zlotnick}(2002)}]{Endres2002}
\bibinfo{author}{\bibfnamefont{D.}~\bibnamefont{Endres}} \bibnamefont{and}
  \bibinfo{author}{\bibfnamefont{A.}~\bibnamefont{Zlotnick}},
  \bibinfo{journal}{Biophys. J.} \textbf{\bibinfo{volume}{83}},
  \bibinfo{pages}{1217} (\bibinfo{year}{2002}).

\bibitem[{\citenamefont{Zlotnick et~al.}(2000)\citenamefont{Zlotnick, Aldrich,
  Johnson, Ceres, and Young}}]{Zlotnick2000}
\bibinfo{author}{\bibfnamefont{A.}~\bibnamefont{Zlotnick}},
  \bibinfo{author}{\bibfnamefont{R.}~\bibnamefont{Aldrich}},
  \bibinfo{author}{\bibfnamefont{J.~M.} \bibnamefont{Johnson}},
  \bibinfo{author}{\bibfnamefont{P.}~\bibnamefont{Ceres}}, \bibnamefont{and}
  \bibinfo{author}{\bibfnamefont{M.~J.} \bibnamefont{Young}},
  \bibinfo{journal}{Virology} \textbf{\bibinfo{volume}{277}},
  \bibinfo{pages}{450} (\bibinfo{year}{2000}).

\bibitem[{\citenamefont{Ceres and Zlotnick}(2002)}]{Ceres2002}
\bibinfo{author}{\bibfnamefont{P.}~\bibnamefont{Ceres}} \bibnamefont{and}
  \bibinfo{author}{\bibfnamefont{A.}~\bibnamefont{Zlotnick}},
  \bibinfo{journal}{Biochemistry} \textbf{\bibinfo{volume}{41}},
  \bibinfo{pages}{11525} (\bibinfo{year}{2002}).

\bibitem[{\citenamefont{Singh and Zlotnick}(2003)}]{Singh2003}
\bibinfo{author}{\bibfnamefont{S.}~\bibnamefont{Singh}} \bibnamefont{and}
  \bibinfo{author}{\bibfnamefont{A.}~\bibnamefont{Zlotnick}},
  \bibinfo{journal}{J. Biol. Chem.} \textbf{\bibinfo{volume}{278}},
  \bibinfo{pages}{18249} (\bibinfo{year}{2003}).

\bibitem[{\citenamefont{Kegel and van~der Schoot}(2004)}]{Kegel2004}
\bibinfo{author}{\bibfnamefont{W.~K.} \bibnamefont{Kegel}} \bibnamefont{and}
  \bibinfo{author}{\bibfnamefont{P.}~\bibnamefont{van~der Schoot}},
  \bibinfo{journal}{Biophys. J.} \textbf{\bibinfo{volume}{86}},
  \bibinfo{pages}{3905} (\bibinfo{year}{2004}).

\bibitem[{\citenamefont{Belyi and Muthukumar}(2006)}]{Belyi2006}
\bibinfo{author}{\bibfnamefont{V.~A.} \bibnamefont{Belyi}} \bibnamefont{and}
  \bibinfo{author}{\bibfnamefont{M.}~\bibnamefont{Muthukumar}},
  \bibinfo{journal}{Proc. Natl. Acad. Sci. U. S. A.}
  \textbf{\bibinfo{volume}{103}}, \bibinfo{pages}{17174}
  (\bibinfo{year}{2006}).

\bibitem[{\citenamefont{Angelescu et~al.}(2006)\citenamefont{Angelescu,
  Bruinsma, and Linse}}]{Angelescu2006}
\bibinfo{author}{\bibfnamefont{D.~G.} \bibnamefont{Angelescu}},
  \bibinfo{author}{\bibfnamefont{R.}~\bibnamefont{Bruinsma}}, \bibnamefont{and}
  \bibinfo{author}{\bibfnamefont{P.}~\bibnamefont{Linse}},
  \bibinfo{journal}{Phys. Rev. E} \textbf{\bibinfo{volume}{73}},
  \bibinfo{pages}{041921} (\bibinfo{year}{2006}).

\bibitem[{\citenamefont{van~der Schoot and Bruinsma}(2005)}]{Schoot2005}
\bibinfo{author}{\bibfnamefont{P.}~\bibnamefont{van~der Schoot}}
  \bibnamefont{and} \bibinfo{author}{\bibfnamefont{R.}~\bibnamefont{Bruinsma}},
  \bibinfo{journal}{Phys. Rev. E} \textbf{\bibinfo{volume}{71}},
  \bibinfo{pages}{061928} (\bibinfo{year}{2005}).

\bibitem[{\citenamefont{Zhang and Glotzer}(2004)}]{Zhang2004}
\bibinfo{author}{\bibfnamefont{Z.~L.} \bibnamefont{Zhang}} \bibnamefont{and}
  \bibinfo{author}{\bibfnamefont{S.~C.} \bibnamefont{Glotzer}},
  \bibinfo{journal}{Nano Lett.} \textbf{\bibinfo{volume}{4}},
  \bibinfo{pages}{1407} (\bibinfo{year}{2004}).

\bibitem[{\citenamefont{Hu et~al.}(2007)\citenamefont{Hu, Zhang, and
  Shklovskii}}]{Hu2007a}
\bibinfo{author}{\bibfnamefont{T.}~\bibnamefont{Hu}},
  \bibinfo{author}{\bibfnamefont{R.}~\bibnamefont{Zhang}}, \bibnamefont{and}
  \bibinfo{author}{\bibfnamefont{B.~I.} \bibnamefont{Shklovskii}},
  \bibinfo{journal}{arXiv:q-bio/0610009v4}  (\bibinfo{year}{2007}).

\bibitem[{\citenamefont{Rudnick and Bruinsma}(2005)}]{Rudnick2005}
\bibinfo{author}{\bibfnamefont{J.}~\bibnamefont{Rudnick}} \bibnamefont{and}
  \bibinfo{author}{\bibfnamefont{R.}~\bibnamefont{Bruinsma}},
  \bibinfo{journal}{Phys. Rev. Lett.} \textbf{\bibinfo{volume}{94}},
  \bibinfo{pages}{038101} (\bibinfo{year}{2005}).

\bibitem[{\citenamefont{Hu and Shklovskii}(2007)}]{Hu2007b}
\bibinfo{author}{\bibfnamefont{T.}~\bibnamefont{Hu}} \bibnamefont{and}
  \bibinfo{author}{\bibfnamefont{B.~I.} \bibnamefont{Shklovskii}},
  \bibinfo{journal}{Phys. Rev. E} \textbf{\bibinfo{volume}{75}},
  \bibinfo{pages}{051901} (\bibinfo{year}{2007}).

\bibitem[{\citenamefont{Elrad and Hagan}(2008)}]{Elrad2008}
\bibinfo{author}{\bibfnamefont{O.~M.} \bibnamefont{Elrad}} \bibnamefont{and}
  \bibinfo{author}{\bibfnamefont{M.~F.} \bibnamefont{Hagan}},
  \bibinfo{journal}{Nano Lett.} \textbf{\bibinfo{volume}{8}},
  \bibinfo{pages}{3850} (\bibinfo{year}{2008}).

\bibitem[{\citenamefont{Hagan}(2008)}]{Hagan2008}
\bibinfo{author}{\bibfnamefont{M.~F.} \bibnamefont{Hagan}},
  \bibinfo{journal}{Phys. Rev. E} \textbf{\bibinfo{volume}{77}},
  \bibinfo{pages}{051904} (\bibinfo{year}{2008}).

\bibitem[{\citenamefont{Hagan and Chandler}(2006)}]{Hagan2006}
\bibinfo{author}{\bibfnamefont{M.~F.} \bibnamefont{Hagan}} \bibnamefont{and}
  \bibinfo{author}{\bibfnamefont{D.}~\bibnamefont{Chandler}},
  \bibinfo{journal}{Biophys. J.} \textbf{\bibinfo{volume}{91}},
  \bibinfo{pages}{42} (\bibinfo{year}{2006}).

\bibitem[{\citenamefont{Rapaport et~al.}(1999)\citenamefont{Rapaport, Johnson,
  and Skolnick}}]{Rapaport1999}
\bibinfo{author}{\bibfnamefont{D.~C.} \bibnamefont{Rapaport}},
  \bibinfo{author}{\bibfnamefont{J.~E.} \bibnamefont{Johnson}},
  \bibnamefont{and} \bibinfo{author}{\bibfnamefont{J.}~\bibnamefont{Skolnick}},
  \bibinfo{journal}{Comput. Phys. Commun.} \textbf{\bibinfo{volume}{122}},
  \bibinfo{pages}{231} (\bibinfo{year}{1999}).

\bibitem[{\citenamefont{Rapaport}(2004)}]{Rapaport2004}
\bibinfo{author}{\bibfnamefont{D.}~\bibnamefont{Rapaport}},
  \bibinfo{journal}{Phys. Rev. E.} \textbf{\bibinfo{volume}{70}},
  \bibinfo{pages}{051905} (\bibinfo{year}{2004}).

\bibitem[{\citenamefont{Rapaport}(2008)}]{Rapaport2008}
\bibinfo{author}{\bibfnamefont{D.}~\bibnamefont{Rapaport}},
  \bibinfo{journal}{Phys. Rev. Lett.} \textbf{\bibinfo{volume}{101}},
  \bibinfo{pages}{186101} (\bibinfo{year}{2008}).

\bibitem[{\citenamefont{Nguyen et~al.}(2007)\citenamefont{Nguyen, Reddy, and
  Brooks}}]{Nguyen2007}
\bibinfo{author}{\bibfnamefont{H.~D.} \bibnamefont{Nguyen}},
  \bibinfo{author}{\bibfnamefont{V.~S.} \bibnamefont{Reddy}}, \bibnamefont{and}
  \bibinfo{author}{\bibfnamefont{C.~L.} \bibnamefont{Brooks}},
  \bibinfo{journal}{Nano Lett.} \textbf{\bibinfo{volume}{7}},
  \bibinfo{pages}{338} (\bibinfo{year}{2007}).

\bibitem[{\citenamefont{Zhang and Schwartz}(2006)}]{Zhang2006}
\bibinfo{author}{\bibfnamefont{T.~Q.} \bibnamefont{Zhang}} \bibnamefont{and}
  \bibinfo{author}{\bibfnamefont{R.}~\bibnamefont{Schwartz}},
  \bibinfo{journal}{Biophys. J.} \textbf{\bibinfo{volume}{90}},
  \bibinfo{pages}{57} (\bibinfo{year}{2006}).

\bibitem[{\citenamefont{Sweeney et~al.}(2008)\citenamefont{Sweeney, Zhang, and
  Schwartz}}]{Sweeney2008}
\bibinfo{author}{\bibfnamefont{B.}~\bibnamefont{Sweeney}},
  \bibinfo{author}{\bibfnamefont{T.}~\bibnamefont{Zhang}}, \bibnamefont{and}
  \bibinfo{author}{\bibfnamefont{R.}~\bibnamefont{Schwartz}},
  \bibinfo{journal}{Biophys. J.} \textbf{\bibinfo{volume}{94}},
  \bibinfo{pages}{772} (\bibinfo{year}{2008}).

\bibitem[{\citenamefont{Schwartz et~al.}(1998)\citenamefont{Schwartz, Shor,
  Prevelige, and Berger}}]{Schwartz1998}
\bibinfo{author}{\bibfnamefont{R.}~\bibnamefont{Schwartz}},
  \bibinfo{author}{\bibfnamefont{P.~W.} \bibnamefont{Shor}},
  \bibinfo{author}{\bibfnamefont{P.~E.} \bibnamefont{Prevelige}},
  \bibnamefont{and} \bibinfo{author}{\bibfnamefont{B.}~\bibnamefont{Berger}},
  \bibinfo{journal}{Biophys. J.} \textbf{\bibinfo{volume}{75}},
  \bibinfo{pages}{2626} (\bibinfo{year}{1998}).

\bibitem[{\citenamefont{Wilber et~al.}(2007)\citenamefont{Wilber, Doye, Louis,
  Noya, Miller, and Wong}}]{Wilber2007}
\bibinfo{author}{\bibfnamefont{A.~W.} \bibnamefont{Wilber}},
  \bibinfo{author}{\bibfnamefont{J.~P.~K.} \bibnamefont{Doye}},
  \bibinfo{author}{\bibfnamefont{A.~A.} \bibnamefont{Louis}},
  \bibinfo{author}{\bibfnamefont{E.~G.} \bibnamefont{Noya}},
  \bibinfo{author}{\bibfnamefont{M.~A.} \bibnamefont{Miller}},
  \bibnamefont{and} \bibinfo{author}{\bibfnamefont{P.}~\bibnamefont{Wong}},
  \bibinfo{journal}{J. Chem. Phys.} \textbf{\bibinfo{volume}{127}}
  (\bibinfo{year}{2007}).

\bibitem[{\citenamefont{Nguyen and Brooks}(2008{\natexlab{a}})}]{Nguyen2008}
\bibinfo{author}{\bibfnamefont{H.}~\bibnamefont{Nguyen}} \bibnamefont{and}
  \bibinfo{author}{\bibfnamefont{C.}~\bibnamefont{Brooks}},
  \bibinfo{journal}{Nano Lett.} \textbf{\bibinfo{volume}{8}},
  \bibinfo{pages}{4574} (\bibinfo{year}{2008}{\natexlab{a}}).

\bibitem[{\citenamefont{Nguyen and Brooks}(2008{\natexlab{b}})}]{Nguyen2008b}
\bibinfo{author}{\bibfnamefont{H.~D.} \bibnamefont{Nguyen}} \bibnamefont{and}
  \bibinfo{author}{\bibfnamefont{C.}~\bibnamefont{Brooks}},
  \bibinfo{journal}{Unpublished}  (\bibinfo{year}{2008}{\natexlab{b}}).

\bibitem[{\citenamefont{Dragnea}(2008{\natexlab{a}})}]{Dragnea2008}
\bibinfo{author}{\bibfnamefont{B.}~\bibnamefont{Dragnea}},
  \bibinfo{journal}{Unpublished}  (\bibinfo{year}{2008}{\natexlab{a}}).

\bibitem[{\citenamefont{Lucas et~al.}(2002)\citenamefont{Lucas, Larson, and
  McPherson}}]{Lucas2002}
\bibinfo{author}{\bibfnamefont{R.~W.} \bibnamefont{Lucas}},
  \bibinfo{author}{\bibfnamefont{S.~B.} \bibnamefont{Larson}},
  \bibnamefont{and}
  \bibinfo{author}{\bibfnamefont{A.}~\bibnamefont{McPherson}},
  \bibinfo{journal}{J. Mol. Biol.} \textbf{\bibinfo{volume}{317}},
  \bibinfo{pages}{95} (\bibinfo{year}{2002}).

\bibitem[{\citenamefont{Chen et~al.}(2008)\citenamefont{Chen, Kao, and
  Dragnea}}]{Chen2008}
\bibinfo{author}{\bibfnamefont{C.}~\bibnamefont{Chen}},
  \bibinfo{author}{\bibfnamefont{C.~C.} \bibnamefont{Kao}}, \bibnamefont{and}
  \bibinfo{author}{\bibfnamefont{B.}~\bibnamefont{Dragnea}},
  \bibinfo{journal}{J. Phys. Chem. A} \textbf{\bibinfo{volume}{112}},
  \bibinfo{pages}{9405} (\bibinfo{year}{2008}).

\bibitem[{\citenamefont{Zlotnick et~al.}(1999)\citenamefont{Zlotnick, Johnson,
  Wingfield, Stahl, and Endres}}]{Zlotnick1999}
\bibinfo{author}{\bibfnamefont{A.}~\bibnamefont{Zlotnick}},
  \bibinfo{author}{\bibfnamefont{J.~M.} \bibnamefont{Johnson}},
  \bibinfo{author}{\bibfnamefont{P.~W.} \bibnamefont{Wingfield}},
  \bibinfo{author}{\bibfnamefont{S.~J.} \bibnamefont{Stahl}}, \bibnamefont{and}
  \bibinfo{author}{\bibfnamefont{D.}~\bibnamefont{Endres}},
  \bibinfo{journal}{Biochemistry} \textbf{\bibinfo{volume}{38}},
  \bibinfo{pages}{14644} (\bibinfo{year}{1999}).

\bibitem[{\citenamefont{Casini et~al.}(2004)\citenamefont{Casini, Graham,
  Heine, Garcea, and Wu}}]{Casini2004}
\bibinfo{author}{\bibfnamefont{G.~L.} \bibnamefont{Casini}},
  \bibinfo{author}{\bibfnamefont{D.}~\bibnamefont{Graham}},
  \bibinfo{author}{\bibfnamefont{D.}~\bibnamefont{Heine}},
  \bibinfo{author}{\bibfnamefont{R.~L.} \bibnamefont{Garcea}},
  \bibnamefont{and} \bibinfo{author}{\bibfnamefont{D.~T.} \bibnamefont{Wu}},
  \bibinfo{journal}{Virology} \textbf{\bibinfo{volume}{325}},
  \bibinfo{pages}{320} (\bibinfo{year}{2004}).

\bibitem[{\citenamefont{Siber and Podgornik}(2007)}]{Siber2007}
\bibinfo{author}{\bibfnamefont{A.}~\bibnamefont{Siber}} \bibnamefont{and}
  \bibinfo{author}{\bibfnamefont{R.}~\bibnamefont{Podgornik}},
  \bibinfo{journal}{Phys. Rev. E} \textbf{\bibinfo{volume}{76}},
  \bibinfo{pages}{061906} (\bibinfo{year}{2007}), ISSN
  \bibinfo{issn}{1539-3755}.

\bibitem[{\citenamefont{Pfeiffer and Hirth}(1974)}]{Pfeiffer1974}
\bibinfo{author}{\bibfnamefont{P.}~\bibnamefont{Pfeiffer}} \bibnamefont{and}
  \bibinfo{author}{\bibfnamefont{L.}~\bibnamefont{Hirth}},
  \bibinfo{journal}{Virology} \textbf{\bibinfo{volume}{61}},
  \bibinfo{pages}{160} (\bibinfo{year}{1974}).

\bibitem[{\citenamefont{Cuillel et~al.}(1983)\citenamefont{Cuillel, Zulauf, and
  Jacrot}}]{Cuillel1983}
\bibinfo{author}{\bibfnamefont{M.}~\bibnamefont{Cuillel}},
  \bibinfo{author}{\bibfnamefont{M.}~\bibnamefont{Zulauf}}, \bibnamefont{and}
  \bibinfo{author}{\bibfnamefont{B.}~\bibnamefont{Jacrot}},
  \bibinfo{journal}{J. Mol. Biol.} \textbf{\bibinfo{volume}{164}},
  \bibinfo{pages}{589} (\bibinfo{year}{1983}).

\bibitem[{\citenamefont{Berthetcolominas
  et~al.}(1987)\citenamefont{Berthetcolominas, Cuillel, Koch, Vachette, and
  Jacrot}}]{Berthetcolominas1987}
\bibinfo{author}{\bibfnamefont{C.}~\bibnamefont{Berthetcolominas}},
  \bibinfo{author}{\bibfnamefont{M.}~\bibnamefont{Cuillel}},
  \bibinfo{author}{\bibfnamefont{M.~H.~J.} \bibnamefont{Koch}},
  \bibinfo{author}{\bibfnamefont{P.}~\bibnamefont{Vachette}}, \bibnamefont{and}
  \bibinfo{author}{\bibfnamefont{B.}~\bibnamefont{Jacrot}},
  \bibinfo{journal}{Eur. Biophys. J. with Biophys. Lett.}
  \textbf{\bibinfo{volume}{15}}, \bibinfo{pages}{159} (\bibinfo{year}{1987}).

\bibitem[{\citenamefont{Siber and Podgornik}(2008)}]{Siber2008}
\bibinfo{author}{\bibfnamefont{A.}~\bibnamefont{Siber}} \bibnamefont{and}
  \bibinfo{author}{\bibfnamefont{R.}~\bibnamefont{Podgornik}},
  \bibinfo{journal}{Phys. Rev. E} \textbf{\bibinfo{volume}{78}},
  \bibinfo{pages}{051915} (\bibinfo{year}{2008}).

\bibitem[{\citenamefont{Scheutjens and Fleer}(1979)}]{Scheutjens1979}
\bibinfo{author}{\bibfnamefont{J.}~\bibnamefont{Scheutjens}} \bibnamefont{and}
  \bibinfo{author}{\bibfnamefont{G.~J.} \bibnamefont{Fleer}},
  \bibinfo{journal}{J. Phys. Chem.} \textbf{\bibinfo{volume}{83}},
  \bibinfo{pages}{1619} (\bibinfo{year}{1979}).

\bibitem[{\citenamefont{Bohmer et~al.}(1990)\citenamefont{Bohmer, Evers, and
  Scheutjens}}]{Bohmer1990}
\bibinfo{author}{\bibfnamefont{M.~R.} \bibnamefont{Bohmer}},
  \bibinfo{author}{\bibfnamefont{O.~A.} \bibnamefont{Evers}}, \bibnamefont{and}
  \bibinfo{author}{\bibfnamefont{J.}~\bibnamefont{Scheutjens}},
  \bibinfo{journal}{Macromolecules} \textbf{\bibinfo{volume}{23}},
  \bibinfo{pages}{2288} (\bibinfo{year}{1990}).

\bibitem[{\citenamefont{Bohmer et~al.}(1991)\citenamefont{Bohmer, Koopal, and
  Lyklema}}]{Bohmer1991}
\bibinfo{author}{\bibfnamefont{M.~R.} \bibnamefont{Bohmer}},
  \bibinfo{author}{\bibfnamefont{L.~K.} \bibnamefont{Koopal}},
  \bibnamefont{and} \bibinfo{author}{\bibfnamefont{J.}~\bibnamefont{Lyklema}},
  \bibinfo{journal}{J. Phys. Chem.} \textbf{\bibinfo{volume}{95}},
  \bibinfo{pages}{9569} (\bibinfo{year}{1991}).

\bibitem[{\citenamefont{Israels et~al.}(1994)\citenamefont{Israels, Leermakers,
  and Fleer}}]{Israels1994}
\bibinfo{author}{\bibfnamefont{R.}~\bibnamefont{Israels}},
  \bibinfo{author}{\bibfnamefont{F.~A.~M.} \bibnamefont{Leermakers}},
  \bibnamefont{and} \bibinfo{author}{\bibfnamefont{G.~J.} \bibnamefont{Fleer}},
  \bibinfo{journal}{Macromolecules} \textbf{\bibinfo{volume}{27}},
  \bibinfo{pages}{3087} (\bibinfo{year}{1994}).

\bibitem[{\citenamefont{Dragnea}(2008{\natexlab{b}})}]{Dragnea2008privcomm}
\bibinfo{author}{\bibfnamefont{B.}~\bibnamefont{Dragnea}},
  \emph{\bibinfo{title}{Personal communication}}
  (\bibinfo{year}{2008}{\natexlab{b}}).

\bibitem[{\citenamefont{Kawakami et~al.}(2006)\citenamefont{Kawakami, Byrne,
  Khatri, McLeish, and Smith}}]{Kawakami2006}
\bibinfo{author}{\bibfnamefont{M.}~\bibnamefont{Kawakami}},
  \bibinfo{author}{\bibfnamefont{K.}~\bibnamefont{Byrne}},
  \bibinfo{author}{\bibfnamefont{B.~S.} \bibnamefont{Khatri}},
  \bibinfo{author}{\bibfnamefont{T.~C.~B.} \bibnamefont{McLeish}},
  \bibnamefont{and} \bibinfo{author}{\bibfnamefont{D.~A.} \bibnamefont{Smith}},
  \bibinfo{journal}{Chemphyschem} \textbf{\bibinfo{volume}{7}},
  \bibinfo{pages}{1710} (\bibinfo{year}{2006}).

\bibitem[{\citenamefont{Oesterhelt et~al.}(1999)\citenamefont{Oesterhelt, Rief,
  and Gaub}}]{Oesterhelt1999}
\bibinfo{author}{\bibfnamefont{F.}~\bibnamefont{Oesterhelt}},
  \bibinfo{author}{\bibfnamefont{M.}~\bibnamefont{Rief}}, \bibnamefont{and}
  \bibinfo{author}{\bibfnamefont{H.~E.} \bibnamefont{Gaub}},
  \bibinfo{journal}{New Journal of Physics} \textbf{\bibinfo{volume}{1}},
  \bibinfo{pages}{6} (\bibinfo{year}{1999}).

\bibitem[{\citenamefont{Kienberger et~al.}(2000)\citenamefont{Kienberger,
  Pastushenko, Kada, Gruber, Riener, Schindler, and
  Hinterdorfer}}]{Kienberger2000}
\bibinfo{author}{\bibfnamefont{F.}~\bibnamefont{Kienberger}},
  \bibinfo{author}{\bibfnamefont{V.~P.} \bibnamefont{Pastushenko}},
  \bibinfo{author}{\bibfnamefont{G.}~\bibnamefont{Kada}},
  \bibinfo{author}{\bibfnamefont{H.~J.} \bibnamefont{Gruber}},
  \bibinfo{author}{\bibfnamefont{C.}~\bibnamefont{Riener}},
  \bibinfo{author}{\bibfnamefont{H.}~\bibnamefont{Schindler}},
  \bibnamefont{and}
  \bibinfo{author}{\bibfnamefont{P.}~\bibnamefont{Hinterdorfer}},
  \bibinfo{journal}{Single Molecules} \textbf{\bibinfo{volume}{1}},
  \bibinfo{pages}{123} (\bibinfo{year}{2000}).

\bibitem[{\citenamefont{Manning}(1969)}]{Manning1969}
\bibinfo{author}{\bibfnamefont{G.~S.} \bibnamefont{Manning}},
  \bibinfo{journal}{J. Chem. Phys.} \textbf{\bibinfo{volume}{51}},
  \bibinfo{pages}{934} (\bibinfo{year}{1969}).

\bibitem[{\citenamefont{Muthukumar}(1996)}]{Muthukumar1996}
\bibinfo{author}{\bibfnamefont{M.}~\bibnamefont{Muthukumar}},
  \bibinfo{journal}{J. Chem. Phys.} \textbf{\bibinfo{volume}{105}},
  \bibinfo{pages}{5183} (\bibinfo{year}{1996}).

\bibitem[{\citenamefont{Patra and Yethiraj}(1999)}]{Patra1999}
\bibinfo{author}{\bibfnamefont{C.~N.} \bibnamefont{Patra}} \bibnamefont{and}
  \bibinfo{author}{\bibfnamefont{A.}~\bibnamefont{Yethiraj}},
  \bibinfo{journal}{J. Phys. Chem. B} \textbf{\bibinfo{volume}{103}},
  \bibinfo{pages}{6080} (\bibinfo{year}{1999}).

\bibitem[{\citenamefont{Kundagrami and Muthukumar}(2008)}]{Kundagrami2008}
\bibinfo{author}{\bibfnamefont{A.}~\bibnamefont{Kundagrami}} \bibnamefont{and}
  \bibinfo{author}{\bibfnamefont{M.}~\bibnamefont{Muthukumar}},
  \bibinfo{journal}{J. Chem. Phys.} \textbf{\bibinfo{volume}{128}},
  \bibinfo{pages}{244901} (\bibinfo{year}{2008}).

\bibitem[{\citenamefont{Gershevitz and Sukenik}(2004)}]{Gershevitz2004}
\bibinfo{author}{\bibfnamefont{O.}~\bibnamefont{Gershevitz}} \bibnamefont{and}
  \bibinfo{author}{\bibfnamefont{C.~N.} \bibnamefont{Sukenik}},
  \bibinfo{journal}{Journal of the American Chemical Society}
  \textbf{\bibinfo{volume}{126}}, \bibinfo{pages}{482} (\bibinfo{year}{2004}).

\bibitem[{\citenamefont{Ceres et~al.}(2004)\citenamefont{Ceres, Stray, and
  Zlotnick}}]{Ceres2004}
\bibinfo{author}{\bibfnamefont{P.}~\bibnamefont{Ceres}},
  \bibinfo{author}{\bibfnamefont{S.~J.} \bibnamefont{Stray}}, \bibnamefont{and}
  \bibinfo{author}{\bibfnamefont{A.}~\bibnamefont{Zlotnick}},
  \bibinfo{journal}{J. Virol.} \textbf{\bibinfo{volume}{78}},
  \bibinfo{pages}{9538} (\bibinfo{year}{2004}).

\bibitem[{\citenamefont{Bruinsma et~al.}(2003)\citenamefont{Bruinsma, Gelbart,
  Reguera, Rudnick, and Zandi}}]{Bruinsma2003}
\bibinfo{author}{\bibfnamefont{R.~F.} \bibnamefont{Bruinsma}},
  \bibinfo{author}{\bibfnamefont{W.~M.} \bibnamefont{Gelbart}},
  \bibinfo{author}{\bibfnamefont{D.}~\bibnamefont{Reguera}},
  \bibinfo{author}{\bibfnamefont{J.}~\bibnamefont{Rudnick}}, \bibnamefont{and}
  \bibinfo{author}{\bibfnamefont{R.}~\bibnamefont{Zandi}},
  \bibinfo{journal}{Phys. Rev. Lett.} \textbf{\bibinfo{volume}{90}},
  \bibinfo{pages}{248101} (\bibinfo{year}{2003}).

\bibitem[{\citenamefont{Berg and Purcell}(1977)}]{Berg1977}
\bibinfo{author}{\bibfnamefont{H.~C.} \bibnamefont{Berg}} \bibnamefont{and}
  \bibinfo{author}{\bibfnamefont{E.~M.} \bibnamefont{Purcell}},
  \bibinfo{journal}{Biophys. J.} \textbf{\bibinfo{volume}{20}},
  \bibinfo{pages}{193} (\bibinfo{year}{1977}).

\bibitem[{kad()}]{kadNote}
\bibinfo{note}{In most cases, changing $\kad$ while fixing the ratio
  $\kad/\SRF$, simply scales the time dependence of the predicted light
  scatter. This scaling does not hold under conditions of: high functionalized
  charge density $\sigc$, high subunit-subunit binding strength $\gb$, and slow
  subunit adsorption (low $\kad$), in which case empty capsid assembly competes
  with subunit adsorption onto nanoparticles.}

\bibitem[{str()}]{strongAcidNote}
\bibinfo{note}{Assembly into well-formed capsids was not observed
  experimentally for surface functionalization with strong acid groups and
  $\sigc=3$, which could be due to imperfect size matching between capsid and
  core sizes or impeded surface diffusion of strongly absorbed subunits.
  However, we still present the full range of charge loading fractions for
  strong acid groups in order to fully explore the influence of surface charge
  density on assembly kinetics.}

\bibitem[{\citenamefont{Aniagyei et~al.}(2009)\citenamefont{Aniagyei, Kennedy,
  Stein, Willits, Douglas, Young, De, Rotello, Srisathiyanarayanan, Kao
  et~al.}}]{Aniagyei2009}
\bibinfo{author}{\bibfnamefont{S.~E.} \bibnamefont{Aniagyei}},
  \bibinfo{author}{\bibfnamefont{C.~J.} \bibnamefont{Kennedy}},
  \bibinfo{author}{\bibfnamefont{B.}~\bibnamefont{Stein}},
  \bibinfo{author}{\bibfnamefont{D.~A.} \bibnamefont{Willits}},
  \bibinfo{author}{\bibfnamefont{T.}~\bibnamefont{Douglas}},
  \bibinfo{author}{\bibfnamefont{M.~J.} \bibnamefont{Young}},
  \bibinfo{author}{\bibfnamefont{M.}~\bibnamefont{De}},
  \bibinfo{author}{\bibfnamefont{V.~M.} \bibnamefont{Rotello}},
  \bibinfo{author}{\bibfnamefont{D.}~\bibnamefont{Srisathiyanarayanan}},
  \bibinfo{author}{\bibfnamefont{C.~C.} \bibnamefont{Kao}},
  \bibnamefont{et~al.}, \bibinfo{journal}{Nano Letters}
  \textbf{\bibinfo{volume}{9}}, \bibinfo{pages}{393} (\bibinfo{year}{2009}).

\bibitem[{\citenamefont{Tang et~al.}(2006)\citenamefont{Tang, Johnson, Dryden,
  Young, Zlotnick, and Johnson}}]{Tang2006}
\bibinfo{author}{\bibfnamefont{J.~H.} \bibnamefont{Tang}},
  \bibinfo{author}{\bibfnamefont{J.~M.} \bibnamefont{Johnson}},
  \bibinfo{author}{\bibfnamefont{K.~A.} \bibnamefont{Dryden}},
  \bibinfo{author}{\bibfnamefont{M.~J.} \bibnamefont{Young}},
  \bibinfo{author}{\bibfnamefont{A.}~\bibnamefont{Zlotnick}}, \bibnamefont{and}
  \bibinfo{author}{\bibfnamefont{J.~E.} \bibnamefont{Johnson}},
  \bibinfo{journal}{J. Struct. Biol.} \textbf{\bibinfo{volume}{154}},
  \bibinfo{pages}{59} (\bibinfo{year}{2006}).

\bibitem[{\citenamefont{Johnson et~al.}(2005)\citenamefont{Johnson, Tang,
  Nyame, Willits, Young, and Zlotnick}}]{Johnson2005}
\bibinfo{author}{\bibfnamefont{J.~M.} \bibnamefont{Johnson}},
  \bibinfo{author}{\bibfnamefont{J.~H.} \bibnamefont{Tang}},
  \bibinfo{author}{\bibfnamefont{Y.}~\bibnamefont{Nyame}},
  \bibinfo{author}{\bibfnamefont{D.}~\bibnamefont{Willits}},
  \bibinfo{author}{\bibfnamefont{M.~J.} \bibnamefont{Young}}, \bibnamefont{and}
  \bibinfo{author}{\bibfnamefont{A.}~\bibnamefont{Zlotnick}},
  \bibinfo{journal}{Nano Lett.} \textbf{\bibinfo{volume}{5}},
  \bibinfo{pages}{765} (\bibinfo{year}{2005}).

\bibitem[{\citenamefont{Endres et~al.}(2005)\citenamefont{Endres, Miyahara,
  Moisant, and Zlotnick}}]{Endres2005}
\bibinfo{author}{\bibfnamefont{D.}~\bibnamefont{Endres}},
  \bibinfo{author}{\bibfnamefont{M.}~\bibnamefont{Miyahara}},
  \bibinfo{author}{\bibfnamefont{P.}~\bibnamefont{Moisant}}, \bibnamefont{and}
  \bibinfo{author}{\bibfnamefont{A.}~\bibnamefont{Zlotnick}},
  \bibinfo{journal}{Protein Sci.} \textbf{\bibinfo{volume}{14}},
  \bibinfo{pages}{1518} (\bibinfo{year}{2005}).

\bibitem[{\citenamefont{Siepmann et~al.}(1992)\citenamefont{Siepmann, McDonald,
  and Frenkel}}]{Siepmann1992}
\bibinfo{author}{\bibfnamefont{J.~I.} \bibnamefont{Siepmann}},
  \bibinfo{author}{\bibfnamefont{I.~R.} \bibnamefont{McDonald}},
  \bibnamefont{and} \bibinfo{author}{\bibfnamefont{D.}~\bibnamefont{Frenkel}},
  \bibinfo{journal}{Journal of Physics-Condensed Matter}
  \textbf{\bibinfo{volume}{4}}, \bibinfo{pages}{679} (\bibinfo{year}{1992}).

\bibitem[{\citenamefont{van Lent and Scheutjens}(1989)}]{Vanlent1989}
\bibinfo{author}{\bibfnamefont{B.}~\bibnamefont{van Lent}} \bibnamefont{and}
  \bibinfo{author}{\bibfnamefont{J.}~\bibnamefont{Scheutjens}},
  \bibinfo{journal}{Macromolecules} \textbf{\bibinfo{volume}{22}},
  \bibinfo{pages}{1931} (\bibinfo{year}{1989}).

\bibitem[{\citenamefont{Cosgrove et~al.}(1987)\citenamefont{Cosgrove, Heath,
  van Lent, Leermakers, and Scheutjens}}]{Cosgrove1987}
\bibinfo{author}{\bibfnamefont{T.}~\bibnamefont{Cosgrove}},
  \bibinfo{author}{\bibfnamefont{T.}~\bibnamefont{Heath}},
  \bibinfo{author}{\bibfnamefont{B.}~\bibnamefont{van Lent}},
  \bibinfo{author}{\bibfnamefont{F.}~\bibnamefont{Leermakers}},
  \bibnamefont{and}
  \bibinfo{author}{\bibfnamefont{J.}~\bibnamefont{Scheutjens}},
  \bibinfo{journal}{Macromolecules} \textbf{\bibinfo{volume}{20}},
  \bibinfo{pages}{1692} (\bibinfo{year}{1987}).

\bibitem[{\citenamefont{Evers et~al.}(1990)\citenamefont{Evers, Scheutjens, and
  Fleer}}]{Evers1990}
\bibinfo{author}{\bibfnamefont{O.~A.} \bibnamefont{Evers}},
  \bibinfo{author}{\bibfnamefont{J.}~\bibnamefont{Scheutjens}},
  \bibnamefont{and} \bibinfo{author}{\bibfnamefont{G.~J.} \bibnamefont{Fleer}},
  \bibinfo{journal}{Macromolecules} \textbf{\bibinfo{volume}{23}},
  \bibinfo{pages}{5221} (\bibinfo{year}{1990}).

\bibitem[{\citenamefont{van Lent et~al.}(1990)\citenamefont{van Lent, Israels,
  Scheutjens, and Fleer}}]{Lent1990}
\bibinfo{author}{\bibfnamefont{B.}~\bibnamefont{van Lent}},
  \bibinfo{author}{\bibfnamefont{R.}~\bibnamefont{Israels}},
  \bibinfo{author}{\bibfnamefont{J.}~\bibnamefont{Scheutjens}},
  \bibnamefont{and} \bibinfo{author}{\bibfnamefont{G.~J.} \bibnamefont{Fleer}},
  \bibinfo{journal}{J. Colloid Interface Sci.} \textbf{\bibinfo{volume}{137}},
  \bibinfo{pages}{380} (\bibinfo{year}{1990}).

\end{thebibliography}

\end{document}